\documentclass{aa}
\usepackage{graphicx}
\usepackage{txfonts}
\usepackage{epsf}  
\usepackage{amssymb}  
\usepackage{rotating}
\usepackage{lscape} 
\usepackage{longtable}
\usepackage{color}
\usepackage{natbib}
\bibpunct{(}{)}{;}{a}{}{,} 
\begin{document}
%
%
 \newcommand{\betacrbbold}{\mbox{\boldmath $\beta$}~CrB} 

 \newcommand{\betacrb}{$\beta$~CrB}
 \newcommand{\tenaql}{10~Aql}
 \newcommand{\gammaequ}{$\gamma$~Equ}

 \newcommand{\thetald}{$\theta_{\rm LD}$}
 \newcommand{\fbol}{$f_{\rm bol}$}
 \newcommand{\acir}{$\alpha$~Cir}
 \newcommand{\ie}{i.e.}
 \newcommand{\eg}{e.g.}
 \newcommand{\cf}{cf.}
 \newcommand{\kms}{km\,s$^{-1}$}
\newcommand{\teff}{\ensuremath{T_{\rm eff}}} 
 \newcommand{\teffii}{T_\rm{eff}}
 \newcommand{\teffsun}{$T_\rm{eff},{\odot}$}
 \newcommand{\logg}{\ensuremath{\log g}}
 \newcommand{\feh}{[Fe/H]}
 \newcommand{\msun}{${\rm M}_{\odot}$}
 \newcommand{\rsun}{${\rm R}_{\odot}$}
 \newcommand{\lsun}{${\rm L}_{\odot}$}
\newcommand{\percent}{\,{\%}}
\newcommand{\kepler}{{\em Kepler}}

\newcommand{\vsini}{\ensuremath{v \sin i}}
\newcommand{\feone}{Fe\,{\sc I}}
\newcommand{\fetwo}{Fe\,{\sc II}}
\newcommand{\kic}{KIC\,8410637}

%


%
\title{\kic: a 408-day period eclipsing binary containing a pulsating giant star.
\thanks{Based on observations made with the Nordic Optical Telescope, operated
on the island of La Palma jointly by Denmark, Finland, Iceland,
Norway, and Sweden, in the Spanish Observatorio del Roque de los
Muchachos of the Instituto de Astrof\'{\i}sica de Canarias.}}
\titlerunning{\kic: a 408-day period eclipsing binary containing a pulsating giant star}
\authorrunning{S. Frandsen et al.}
\author{
S.~Frandsen\inst{1,11}
\and 
H.~Lehmann\inst{2}
\and
S.~Hekker\inst{3}
\and
J.~Southworth\inst{4}
\and
J.~Debosscher\inst{5}
\and
P.~Beck\inst{5}
\and
M.~Hartmann\inst{2}
\and
A.~Pigulski\inst{6}
\and
G.~Kopacki\inst{6}
\and
Z.~Ko{\l}aczkowski\inst{6}
\and
M.~St\c{e}\'slicki\inst{6,9}
\and
A. O.~Thygesen\inst{1,10} 
\and
K.~Brogaard\inst{1,7}
\and
Y.~Elsworth\inst{8}
} 
\offprints{S.~Frandsen}
\mail{srf@phys.au.dk}
\institute{Department of Physics and Astronomy, Aarhus University, DK-8000 Aarhus C, Denmark
\and Th\"uringer Landessternwarte Tautenburg, Sternwarte 5, D-07778 Tautenburg, Germany
\and Astronomical Institute `Anton Pannekoek', University of Amsterdam, Science Park 904, 1098 HX, Amsterdam, The Netherlands
\and Astrophysics Group, Keele University, Staffordshire, ST5 5BG, UK
\and Instituut voor Sterrenkunde, K.U.\ Leuven, Celestijnenlaan 200D, 3001 Leuven, Belgium
\and Instytut Astronomiczny Uniwersytetu Wroc{\l}awskiego, Kopernika 11, 51-622 Wroc{\l}aw, Poland
\and Department of Physics and Astronomy, University of Victoria, P.O.\ Box 3055, Victoria, B.C., V8W3P6, Canada
\and School of Physics and Astronomy, University of Birmingham, Edgbaston, Birmingham, B15 2TT, UK
\and Space Research Centre, Polish Academy of Sciences, Solar Physics Division, Kopernika 11, 51-622 Wroc{\l}aw, Poland
\and Zentrum f\"ur Astronomie der Universit\"at Heidelberg, Landessternwarte, K\"onigstuhl 12, 69117 Heidelberg, Germany
\and Stellar Astrophysics Centre, Department of Physics and Astronomy, Aarhus University, Ny Munkegade 120, DK-8000 Aarhus C, Denmark
}
\date{Received 02-05 2012 ; Accepted 20-06 2013 }
\abstract
{Detached eclipsing binaries (dEBs) are ideal targets for accurately measuring of the masses and radii of their component stars. If at least one of the stars has evolved off the main sequence (MS), the masses and radii give a strict constraint on the age of the stars. Several dEBs containing a bright K giant and a fainter MS star have been discovered by the \kepler\ satellite. The mass and radius of a red giant (RG) star can also be derived from its asteroseismic signal. The parameters determined in this way depend on stellar models and may contain systematic errors. 
It is important to validate the asteroseismically determined mass and radius with independent methods. This can be done when stars are members of stellar clusters or members of dEBs.
}
{ This paper presents an analysis of the dEB system \kic, which consists of an RG and an MS star. The aim is to derive accurate masses and radii for both components and provide the foundation for a strong test of the asteroseismic method and the accuracy of the deduced mass, radius, and age.}
{We analysed high-resolution, high-signal-to-noise spectra from three different spectrographs. We also calculated a fit to the \kepler\ light curve and used ground-based photometry to determine the flux ratios between the component stars in the $BVRI$ passbands.}
{We measured the masses and radii of the stars in the dEB, and the classical parameters \teff, \logg\ ,and \feh\ from the spectra and ground-based photometry. The RG component of \kic\ is most likely in the core helium-burning red clump phase of evolution and has an age and composition that are very similar to the stars in the open cluster NGC\,6819. The mass of the RG in \kic\ should therefore be similar to the mass of RGs in NGC\,6819, thus lending support to the latest version of the asteroseismic scaling relations. This is the first direct measurement of both mass and radius for an RG to be compared with values for RGs from asteroseismic scaling relations thereby providing an accurate comparison. We find excellent agreement between \logg\ values derived from the binary analysis and asteroseismic scaling relations.
}
{We have determined the masses and radii of the two stars in the binary accurately. A detailed asteroseismic analysis will be presented in a forthcoming paper, allowing an informative comparison between the parameters determined for the dEB and from asteroseismology.}
\maketitle
%
%
%

\section{Introduction \label{sec:intro}}

From 2009 to 2013 the NASA \kepler\ mission continuously observed the same field in the sky and measured the flux for thousands of stars. A remarkable advance has been made possible in the exploration of red giant (RG) stars, where stochastically excited oscillations have been observed in more than 10,000 stars by \kepler\ \citep{redgiants}. \citet{stello} present an analysis of 13,000 RGs. The asteroseismic data offer the possibility to classify RG stars as red giant branch (RGB) stars or red clump (RC) stars \citep{bedding1} or to study the differential rotation \citep{beck}. This is just the beginning and detailed modelling, as presented by e.g. \citet{chen}, is expected to teach us a lot about the precise structure of the interior of these evolved stars, when applied to a range of RG stars.

It is important to verify how accurate the seismic results are; therefore, we want to measure as many non-seismic parameters as possible with high precision to ensure a coherent picture of the star. This is why we need the classical parameters obtained by photometry \citep{uytterhoeven} and spectroscopy \citep{thygesen}. 

A rare chance for a very accurate test of the results for mass $M$ and radius $R$ from the asteroseismic data (see e.g.\ \citet{kallinger}) is presented by the detection of a few eclipsing binaries (EBs) with an RG component. The first such system is the Kepler target \kic\ that was identified by \citet{hekker}), when only one eclipse was detected. Since then, the list has been growing with the recent publication of about ten new candidates RGs in EBs \citep{gaulme}). The present work aims at establishing such a case using extensive ground-based photometry and spectroscopy. We spectroscopically observed three targets initially (KIC\,5640750, \kic\ and KIC\,9540226), when the periods were not yet known. Based on the inspection of the first reduced spectra we chose \kic\ as the most promising system. The main reason was that \kic\ is the brightest target so we got the most accurate radial velocities. The stellar parameters for \kic\ are derived from photometry and spectroscopy as a baseline for the asteroseismic study. But we are also able to test stellar evolution as we get the mass of the RG, which is closely related to the age of the system.

Detached eclipsing binary systems (dEBs) are known to offer the opportunity to get accurate masses and radii for the two components. \citet{torres} and a long series of papers by Clausen et al.\ (e.g.\ \citealt{bkpeg}) demonstrate that both parameters in favourable cases can be determined with accuracies of the order of 1\%. For only slightly evolved or MS stars with radius and mass close to 1\,\rsun\ and 1\,\msun, systems can be considered detached if the orbital period $P > 1$\,d. For shorter periods the stars interact strongly, and mass transfer occurs during the evolution. Most systems analysed have periods of several days. Due to their large radii, dEBs with an RG component with a mass of the order of 1\msun need to have $P > 30$\,d in order to be detached systems, in particular since the ellipticity of the orbit can be quite large. In the catalogue of EBs by \citet{ebcat} there are 115 EBs with $P > 30$ days. Only three systems, KL\,Cep ($P=256.1$\,d V472\,Sco ($P=208.75$\,d), and RR\,Ari ($P = 47.9$\,d), have a component of spectral type K, probably an RG star. The nature of the secondary components is not known.

EBs with such long periods are difficult to detect. They may only have one or two eclipses each year. Only extensive long-term surveys like OGLE II/III (6833 EBs, 3031 in the LMC, \citet{ogle}) or the TrES project (773 EBs, \citet{trescat}) are likely to find candidates. The OGLE targets are distant and difficult to observe, and the TrES list of EBs has a maximum orbital period of 29 days. A large number of distant EBs have been found by the MACHO project \citep{macho}.
A literature search reveals only one system, TZ For, with $P > 30$\,d and well-understood properties \citep{tzfor}. It consists of a G8\,III and an F7\,IV star in a 76\,d orbit and with masses 1.95\msun\ and 2.05\msun, respectively. Whilst somewhat evolved, these stars are not yet on the RGB.

\kepler\ is the perfect tool for a search for all sorts of dEB systems. The long cadence observations give one data point per 29.4 minutes and the observations extend for years. The number of EBs discovered in the field initially consisted of 2176 and now includes 2708 entries in the \kepler\ EB catalogue\footnote{Latest catalogue at http://archive.stsci.edu/kepler/\-eclip\-sing\_bin\-aries.html} (\citealt{ebcat1} and \citealt{ebcat2}). Only two of the listed EBs are RG stars having periods longer than 30 days: KIC\,9540226 ($P=175.4$\,d) and \kic\ ($P=408.3$\,d); the latter is the target of this paper. The third system, KIC\,5640750, is not included in the catalogue. Two total eclipses has been observed in a data set that includes quarter 13 (Q13). Thus, KIC\,5640750 has a very long period around 3.6\,years. \citet{gaulme} provide a list of nine candidates ($41\,{\rm d} < P < 235\,{rm d}$) from the same catalogue, but spectra are needed to verify the classification. 

Our selected target, \kic, has a \kepler\ magnitude $K_p = 10.77$. The \kepler\ light curve up to quarter 11 (Q11) covers three primary and three secondary eclipses.
From \citet{hekker} we already have a first set of parameters for the RG star: \teff$_{\rm ,RG} = 4680\pm150$\,K and \logg$_{\rm RG} = 2.8\pm0.3$. Analysing the solar-like oscillations yielded a mass $M_{\rm RG} = 1.7\pm0.3$\,\msun\ and a radius $R_{\rm RG} = 11.8\pm0.6$\,\rsun. Some estimates of the parameters of the companion star, which appears to be on the MS, were calculated, even though the secondary eclipse had not yet been observed and the orbital period was unknown. They were \teff$_{\rm ,MS} = 6700\pm200$\,K, \logg$_{\rm MS} = 4.2\pm0.1$ and for the mass and radius $M_{\rm MS} = 1.44\pm0.05$\,\msun\ and $R_{\rm MS} = 1.7\pm0.1$\,\rsun.
\kic\ is also included in the list of EBs analysed by \citet{gaulme}, who gave the following estimates for mass and radius: $M_{\rm MS} = 1.8\pm0.7$\,\msun, $R_{\rm MS} = 1.6\pm0.1$\,\rsun\ ,$M_{\rm RG} = 1.6\pm0.2$\,\msun, and $R_{\rm RG} = 11.0\pm0.5$\,\rsun.
The mass of the MS star is higher than the RG, which is against expectations and later shown here not to be the case.

We now present a detailed analysis of \kic, ordered according to dependencies within the parameters we derive. First we describe the spectroscopic and photometric observations in Sects.\,\ref{sect:hrs} and \ref{sect:phot}. Sect.\,\ref{sect:hrs} provides the basis for the radial velocity measurements (Sect.\,\ref{sect:rv}) and spectrum analysis (Sect.\,\ref{sect:abund}). Sects.\,\ref{sect:orbit} and \ref{sect:lca} describe the derivation of orbital and stellar parameters using the combined spectroscopic and photometric data. Sect.\,\ref{sect:dist} considers the measured distance to the system. Finally, the combined results are discussed in Sect.\,\ref{sect:disc} and conclusions are given in Sect.\,\ref{sect:concl}.

{\it Terminology.}
The standard distinction between primary and secondary eclipses is that the former is deeper than the latter (e.g.\ \citealt{Hilditch01book}). By definition, the primary star is the one eclipsed during primary eclipse, which in turn means that it is the hotter of the two stars. In the case of \kic\ ,this definition means that the dwarf is the primary star and the giant is the secondary star. This is somewhat counterintuitive and has the potential to cause confusion. In the work we therefore avoid using the phrases `primary star' and `secondary star', in favour of the explicit `red giant star' (RG) and `main sequence star' (MS) alternatives.

\begin{figure*}
\centering
\includegraphics[width=\textwidth]{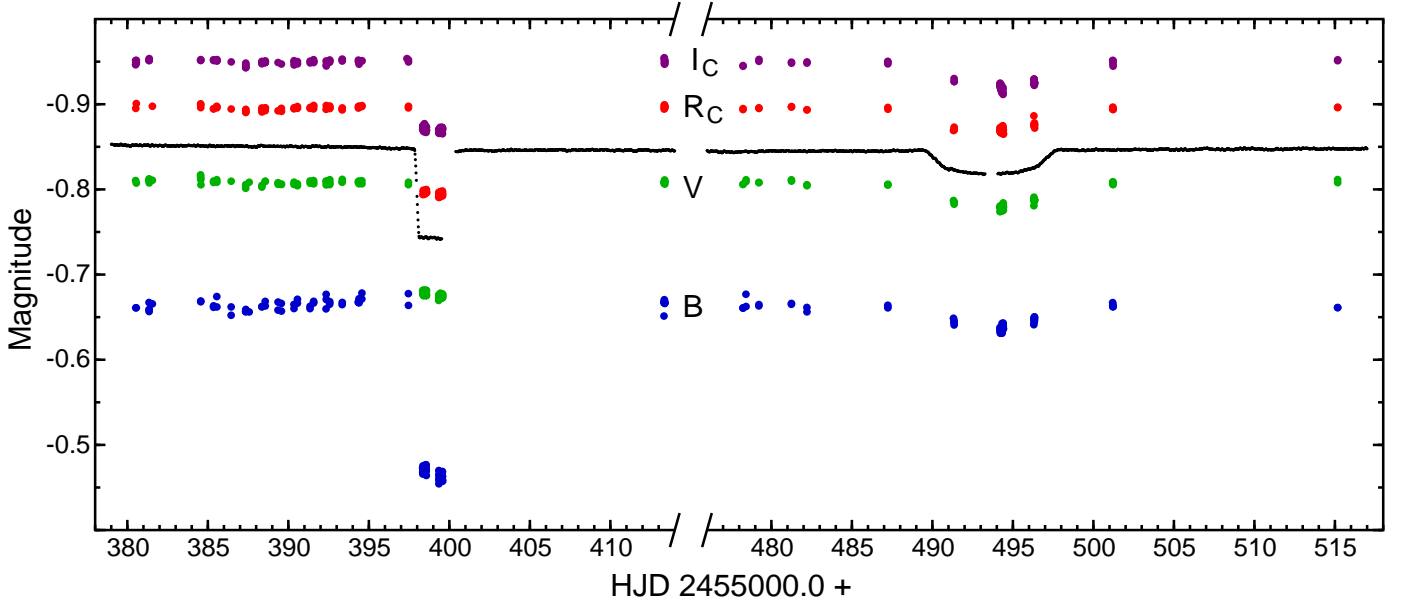}
\caption{Differential Bia{\l}k\'ow $BV(RI)_{\rm C}$ light curves of \kic\ in the vicinity of the 
primary (July 2010) and secondary (October 2010) eclipses. For reference, the \kepler\ light curve, 
shifted arbitrarily in magnitude, is shown. Note the differences in the depth of primary eclipse.}
\label{kic-bia}
\end{figure*}

\section{Ground-based observations \label{sect:obs}}

The long period of \kic\ required ground-based observations over several observing seasons to achieve complete phase coverage.
\begin{table}
\centering
\caption{Summary of the spectroscopic observations.}
\begin{tabular}{llccc}
\hline\hline
Telescope & Instrument & $R$ & $N$   & S/N (per pixel) \\
\hline
NOT  2.5m         & FIES   & 67\,000 & 15 & 43 \\
TLS  2.0m         & CES    & 32\,000 & 12 & 47 \\
\textsc{Mercator}  1.2m & HERMES & 86\,000 & 16 & 34 \\
\hline
\end{tabular}
\label{obstable}
\end{table}

\subsection{High-resolution spectroscopy} \label{sect:hrs}

Three different telescopes were used to obtain spectroscopic data of \kic. The observations are summarised in Table\,\ref{obstable}, listing for each instrument the resolving power ($R$), the number of spectra obtained ($N$), and the typical signal-to-noise ratio (S/N). Altogether we have 43 spectra obtained over a period of 632 days.

At the Nordic Optical Telescope (NOT) we used the FIES spectrograph in high-resolution mode\footnote{http://www.not.iac.es/instruments/fies/}. The exposures were 30\,min long to obtain sufficient S/N whilst avoiding problems with cosmic rays. The data were reduced using the {\sc FIEStool} package and wavelength calibrations were obtained using ThAr spectra framing the target exposures.

The \textsc{Mercator} telescope was used with the high-resolution HERMES spectrograph\footnote{http://fys.kuleuven.be/ster/instruments/hermes} \citep{hermes}. The exposure times were mostly 30\,min with a few shorter or longer exposures. The spectra were reduced with the instrument-specific pipeline \citep{hermes}. ThAr calibration frames were taken at the beginning and end of each observing night.  The pipeline performs all standard corrections and merges the \'echelle orders to a final one-dimensional spectrum.

At the Th\"uringer Landes Sternwarte (TLS) the CES instrument\footnote{http://www.tls-tautenburg.de/TLS/index.php?id=58\&L=1} was used. These observations had exposure times of 40\,min, except for one shorter exposure interrupted by clouds. Data reduction was performed with standard {\sc eso-midas} packages, and corrections for instrumental shifts were applied by using a large number of telluric O$_2$ lines. 

\begin{table}%
\centering
\caption{Flux ratios in the $BVRI$ filters.}
\begin{tabular}{lcr}
\hline\hline
Filter & $ F_{\rm RG}/F_{\rm MS}$ & $\sigma_{\rm std}$ \\
\hline
$B$        & ~~5.03 & 0.08 \\
$V$        & ~~7.86 &  0.11 \\
$R_{\rm C}$ & 10.37 & 0.06 \\
$I_{\rm C}$ & 13.49 & 0.29 \\
\hline
\end{tabular}
\label{frat}
\end{table}

\subsection{Ground-based photometry} \label{sect:phot}

$BV(RI)_{\rm C}$ ground-based photometry of \kic\ was taken at the Bia{\l}k\'ow station, University of Wroc{\l}aw, with a 60\,cm reflecting telescope equipped with an Andor Tech.~DW\,432 BV back-illuminated CCD camera. The observations started on 2010 July 2. At that time the orbital period of the system was not known and only a single primary and a single secondary eclipse had been observed by \kepler. Fortunately, the drop of brightness of \kic\ due to the second primary eclipse was detected on two nights, July 20/21 and 21/22, 2010, allowing for the first estimation of the orbital period ($408.8 \pm 0.5$\,d). Accordingly, observations around the secondary eclipse were scheduled three months later, in October and November 2010. In total, $BV(RI)_{\rm C}$ data were taken during 27 observing nights. They were calibrated in a standard way and reduced by means of the {\sc daophot} package \citep{stet87}. 

Differential magnitudes were calculated with respect to six relatively bright nearby stars, and are shown in Fig.\,\ref{kic-bia}. Flux ratios between the two components were calculated as part of the light curve analysis (see Sect.\,\ref{sect:lca}), adopting the geometry of the system as determined from the \kepler\ data (Table\,\ref{frat}). It can be seen that due to the large difference in effective temperatures of the components, the depth of the primary eclipse is strongly dependent on wavelength (Fig.\,\ref{kic-bia}).

The flux ratios given in Table\,\ref{frat} can be used to interpolate the flux ratio $f(\lambda)$ as function of wavelength by the following approximation:
\begin{equation} \label{fratio}
f(\lambda)^{-1} = -4.1523 + 2.1778 \times 10^{-3} \lambda
\end{equation}
where $\lambda$ is in \AA. The MS component is bluer than the RG and correspondingly the flux ratio is highest in the $B$-band, where $f \sim 0.20$, decreasing to a small value $f \sim 0.05$ in the $I$-band. Therefore, the blue part of the spectrum is where the MS star is most prominent. This flux ratio is needed as input to the {\sc todcor} and {\sc korel} reduction methods discussed later.


The photometric flux ratios are confirmed by an analysis of the observed combined RG and MS spectrum. 
For each spectral order a flux ratio is derived minimising the difference between the observed, normalised
spectrum and a synthetic spectrum consisting of a combination of template spectra for the RG and MS star.
The parameters for the template spectra are derived in the following sections. 
We find the same trend as given by Eq.\ \ref{fratio}, but the spectroscopic results are quite sensitive
to the broadening parameters $v_{\rm macro}$ and \vsini. Therefore we later use Eq.\ \ref{fratio}, which tells us
that, despite the lower overall flux, the blue orders are better for measuring the velocities of the MS star.

\section{Radial velocities \label{sect:rv}}

The radial velocity (RV) of the RG can be determined rather easily from a spectrum even when the exposure time is short. The MS star, however, only contributes typically 10\% of the total light. Due to the long orbital period, the RV separation of the two stars is often quite small and the RV changes quite slowly. The main difficulty is therefore to measure the RV of the fainter MS object.
We will use three alternative approaches to measure the RVs of the two stars.

\subsection{Using line broadening functions}

The line broadening function (LBF) technique assumes that a function can be computed that, convolved with a proper template, can reproduce the
observed spectrum. 
Using a single template spectrum LBFs were computed that, convolved with a template, give optimal fits to the observed spectra at any given epoch. The RVs were then derived by fitting a model function to the LBF for each spectrum. When the RV difference is small, the two peaks in the LBF overlap and are difficult to fit simultaneously. The technique was developed and described by \citet{rucinski3,rucinski2,rucinski1}. It has been used with success by e.g.\ \citet{bkpeg} for the EB BK Peg and for faint dEBs in the open cluster NGC\,6791 (\citealt{ngc6791_1} and \citealt{ngc6791_2}). For \kic\ we chose as template a spectrum of Arcturus, which is a good template for the RG, but far from optimal for the MS star.
However, if one is only interested in the RVs, the non optimal template is not a problem for the MS star.

The method was applied order by order for the orders with sufficient S/N. This order set was chosen individually for each spectrograph by visual inspection. If the extracted \'echelle orders delivered from a spectrograph reduction pipeline were already merged into single spectra, they were divided again into chunks of 10\,nm, which were used as orders. For each spectral range an LBF was computed and at same time an average LBF was generated. Weights were derived by calculating the deviation of each individual LBF from the average LBF and then repeating the calculation of the average LBF. These weights can be used for deriving the average RVs over the range of orders. The weights will diminish the effect of orders which produce an atypical LBF. The reason for that could be the presence of weak and varying telluric lines in the red orders, or interstellar lines in the sodium doublet if that order of the spectrum is included.

\begin{figure}
\includegraphics[width=9cm]{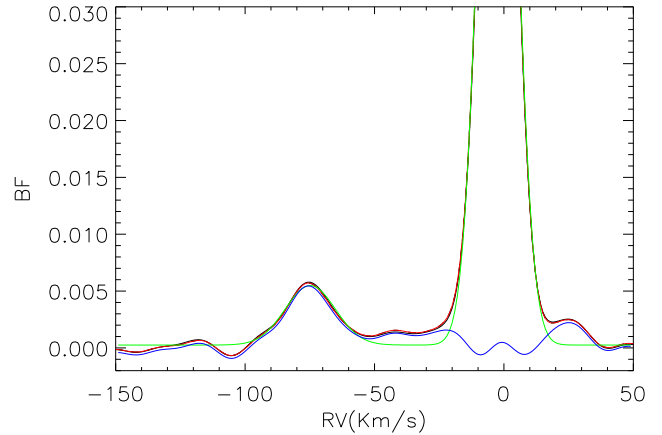}%
\caption{The line broadening function for an epoch with large RV separation. The red 
curve is the LBF derived from the data. The green curve is a fit of two Gaussians. 
The large out-of-range peak comes from the RG and the small peak from the MS star. 
The blue curve is what remains after subtracting a Gaussian function for the RG star.}
\label{lbfs}
\end{figure}

\begin{figure}
\includegraphics[width=9cm]{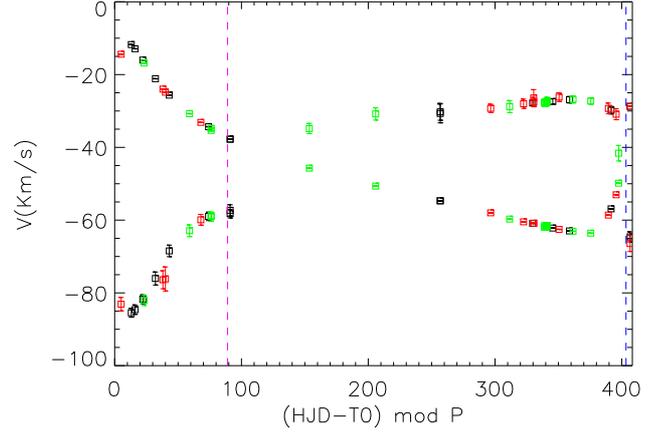}%
\caption{Radial velocities derived with the LBF method. 
Error bars are indicated, but are much smaller than the point sizes
for the RG. Black are observations with the NOT, red from the TLS and green from the 
\textsc{Mercator} telescope. The vertical dashed lines indicate the mid-time of the eclipses.}
\label{rvplot1}
\end{figure}

The function used to fit the LBF was a double Gaussian with a constant background. The background level was determined by the average value at RVs away from the position of the peaks.  Central position, width, and amplitude were free parameters for both peaks for each LBF. Rotational broadening in the model function could be included, but it was verified visually that a Gaussian fits the peak quite well. Fig.\,\ref{lbfs} shows an example where the vertical scale is chosen to show the small MS peak in detail. The background noise is large enough to make the MS peak look asymmetric. There are other epochs with more symmetric MS peaks.
It is clear that it is easy to fit the large peak from the RG precisely, but that the position of the MS peak is harder to determine, when the peaks are not separated. This is evident in Fig.\,\ref{rvplot1}, where points in the range $t = 150-250$\,days for the MS star are systematically off, which becomes particularly clear when trying to fit an orbital solution to the RVs (see later).

The results from the LBF method are listed in Table\,\ref{lbftable}. Times have been converted from heliocentric to barycentric Julian dates as described by \citet{bjd_tdb} in order to be consistent with the time used by \kepler. Only the two last NOT spectra are within an eclipse. They fall almost at the midpoint of a secondary eclipse. As the RG is slowly rotating ($\vsini \sim 2.0$\,\kms, see later) this is not expected to have a significant effect on its measured RVs. 

\begin{table}%
\caption{Radial velocities derived with the LBF method.}
\centering
\begin{tabular}{lrr}
\hline\hline
BJD\_TDB 2455000+  & $v_{\rm RG}$ (\kms) & $v_{\rm MS}$ (\kms) \\
\hline
NOT:\\
660.71310 & $ -54.69\pm  0.05$  & $  -30.68\pm  2.59$ \\
660.73620 & $ -54.69\pm  0.04$  & $  -30.23\pm  2.31$ \\
733.62052 & $ -60.91\pm  0.05$  & $  -27.79\pm  0.76$ \\
749.51187 & $ -62.18\pm  0.05$  & $  -27.39\pm  0.74$ \\
762.64343 & $ -62.98\pm  0.06$  & $  -26.92\pm  0.88$ \\
795.49906 & $ -56.90\pm  0.05$  & $  -29.78\pm  1.04$ \\
810.47562 & $ -28.70\pm  0.05$  & $  -64.60\pm  1.38$ \\
825.34778 & $ -11.75\pm  0.05$  & $  -85.47\pm  1.21$ \\
828.34168 & $ -12.90\pm  0.05$  & $  -84.60\pm  1.30$ \\
834.42851 & $ -16.01\pm  0.06$  & $  -81.75\pm  1.21$ \\
844.39656 & $ -21.13\pm  0.05$  & $  -76.04\pm  1.79$ \\
855.33976 & $ -25.64\pm  0.05$  & $  -68.46\pm  1.59$ \\
886.30030 & $ -34.33\pm  0.05$  & $  -58.98\pm  1.15$ \\
903.31745 & $ -37.68\pm  0.06$  & $  -58.22\pm  1.18$ \\
903.34440 & $ -37.80\pm  0.07$  & $  -57.31\pm  1.55$ \\
TLS:\\       
700.49983 & $ -58.00\pm  0.05$  & $  -29.28\pm  1.14$ \\
726.46415 & $ -60.47\pm  0.06$  & $  -28.01\pm  1.27$ \\
734.42301 & $ -60.87\pm  0.06$  & $  -26.41\pm  2.28$ \\
734.50693 & $ -60.84\pm  0.07$  & $  -27.56\pm  1.29$ \\
754.43702 & $ -62.55\pm  0.06$  & $  -26.19\pm  1.19$ \\
793.35354 & $ -58.62\pm  0.05$  & $  -29.30\pm  1.48$ \\
799.48010 & $ -53.04\pm  0.10$  & $  -30.89\pm  1.52$ \\
810.44951 & $ -28.76\pm  0.06$  & $  -66.44\pm  2.19$ \\
817.33528 & $ -14.40\pm  0.08$  & $  -83.10\pm  1.85$ \\
850.30235 & $ -23.98\pm  0.06$  & $  -76.39\pm  2.48$ \\
852.27125 & $ -24.85\pm  0.06$  & $  -76.17\pm  3.31$ \\
880.20399 & $ -33.14\pm  0.05$  & $  -59.91\pm  1.48$ \\
\textsc{Mercator}:\\       
334.53416 & $ -61.94\pm  0.04$  & $  -27.85\pm  0.82$ \\
334.67372 & $ -61.53\pm  0.05$  & $  -27.59\pm  0.84$ \\
335.57917 & $ -61.90\pm  0.04$  & $  -27.90\pm  0.74$ \\
335.69667 & $ -61.58\pm  0.05$  & $  -27.48\pm  0.88$ \\
336.56763 & $ -61.88\pm  0.05$  & $  -27.69\pm  0.78$ \\
336.72905 & $ -61.47\pm  0.05$  & $  -27.26\pm  1.04$ \\
609.75532 & $ -50.64\pm  0.04$  & $  -30.81\pm  1.69$ \\
715.46097 & $ -59.72\pm  0.05$  & $  -28.80\pm  1.68$ \\
765.44365 & $ -63.07\pm  0.05$  & $  -26.82\pm  0.98$ \\
779.44832 & $ -63.58\pm  0.04$  & $  -27.28\pm  0.88$ \\
801.44158 & $ -49.85\pm  0.05$  & $  -41.61\pm  2.13$ \\
835.38024 & $ -16.82\pm  0.05$  & $  -82.05\pm  1.41$ \\
871.30830 & $ -30.75\pm  0.05$  & $  -62.92\pm  1.61$ \\
888.30934 & $ -35.35\pm  0.05$  & $  -58.77\pm  1.00$ \\
888.32728 & $ -34.80\pm  0.05$  & $  -59.05\pm  1.21$ \\
965.77232 & $ -45.72\pm  0.05$  & $  -34.82\pm  1.41$ \\
\hline
\end{tabular}

\label{lbftable}
\end{table}

\subsection{Using TODCOR} \label{todcor}

To account for the fact that we are dealing with two stars of different spectral types, we have applied the two dimensional cross-correlation technique {\sc todcor} (\citealt{todcor1}, \citealt{todcor2} and \citealt{todcor3}). The spectra used as templates were model spectra. For the RG, a synthetic spectrum corresponding to the best fit described in Sect.\,\ref{sect:abund} was used. The template for the MS star was calculated with parameters $\teff = 6200$\,K, $\logg = 4.0$ and \feh\ $= 0.1$ based on the results in Sect.\,\ref{sect:phot} and Sect.\,\ref{disentangle}. The template spectra were then rotationally broadened using $\vsini = 2.0$\,\kms\ for the RG and $\vsini = 14$\,\kms\ for the MS star.

{\sc todcor} can calculate the flux ratio, but this led to clearly incorrect correlation functions. A better and more stable solution is obtained by the use of the flux ratio from the fit to the $BVRI$ photometry given in Eq.\,\ref{fratio}. As for the LBF method, the spectrum was cut into sections and {\sc todcor} was applied section by section. This permits one to discard sections, where the results deviate significantly from those obtained on average. Weights were derived and used as described in the previous subsection on the LBF method. Alternatively, {\sc todcor} was also applied to the range of the merged spectrum with best S/N and an average correlation function was calculated.

The results are presented in Table\,\ref{todcortable}, where the columns present RVs derived using the following principles:
\begin{itemize}
\item{}
Line broadening function results as a reference (column LBF in Table\,\ref{todcortable}).
\item{}
A weighted {\sc todcor} RV from weights applied to the RV found for each wavelength section. The weights are derived from comparing individual correlation functions to the mean correlation function for all sections (column $TC_{\rm wgh}$).
\item{}
The median of RVs determined for each section (column $TC_{\rm med}$).
\item{}
The RV at the maximum of the mean correlation function (at a resolution of 1\,\kms, column $TC_{\rm max}$).
\item{}
The RV derived from the correlation function derived from the merged spectrum of all orders 
(column $TC_{\rm mer}$).
\end{itemize}
Only results for the NOT are shown in Table\,\ref{todcortable}; RVs for the TLS and Mercator spectra show the same pattern. Trustworthy results are found when the RV difference is large, in which case most methods lead to the same RV with a small spread. Obvious outliers in the RV of the MS star occur when the two stars have a small RV separation. None of the methods are superior at all epochs. Therefore, the final RV for the MS star at any given epoch was taken to be a robust mean of the columns in the table and a weight was calculated that is high when good agreement is found, and low when the column values disagree. These weights were then used in the determination of the orbital parameters.

\begin{table}
\caption{Radial velocities of the MS star derived with {\sc todcor} and 
LBF. The velocities are not corrected for barycentric motion and are only 
shown for spectra from the NOT. The columns are described in the text.}
\centering
\begin{tabular}{llllll}
\hline\hline
  BJD\_TDB    &       LBF  & $TC_{\rm wgh}$ & $TC_{\rm med}$ & $TC_{\rm max}$  & $TC_{\rm mer}$ \\
 2455000+  &  \kms & \kms & \kms & \kms & \kms \\
\hline
 660.71310 & 42.66 & 46.63 & 43.99 & 53.0 & 51.61\\
 660.73620 & 42.24 & 46.62 & 43.07 & 53.0 & 50.60\\
 733.62052 & 32.29 & 34.49 & 33.90 & 34.0 & 37.52\\
 749.51187 & 29.09 & 35.25 & 29.45 & 30.0 & 34.16\\
 762.64343 & 25.49 & 24.35 & 24.98 & 24.0 & 27.25\\
 795.49906 & 22.75 & 28.17 & 26.07 & 30.0 & 31.33\\
 810.47562 & 55.52 & 58.35 & 57.29 & 24.0 & 56.02\\
 825.34778 & 74.93 & 74.92 & 74.73 & 76.0 & 76.69\\
 828.34168 & 73.81 & 73.45 & 72.50 & 74.0 & 73.84\\
 834.42851 & 70.49 & 70.35 & 70.23 & 70.0 & 69.45\\
 844.39656 & 64.33 & 64.68 & 64.62 & 66.0 & 63.58\\
 855.33976 & 56.71 & 58.74 & 59.55 & 59.0 & 57.69\\
 886.30030 & 49.25 & 41.20 & 47.91 & 40.0 & 47.03\\
 903.31745 & 50.85 & 38.46 & 41.23 & 43.0 & 43.35\\
 903.34440 & 49.95 & 51.77 & 44.96 & 35.0 & 49.98\\
\hline
\end{tabular}
\label{todcortable}
\end{table}

\begin{figure*}
\includegraphics[angle=-90, width=18.3cm]{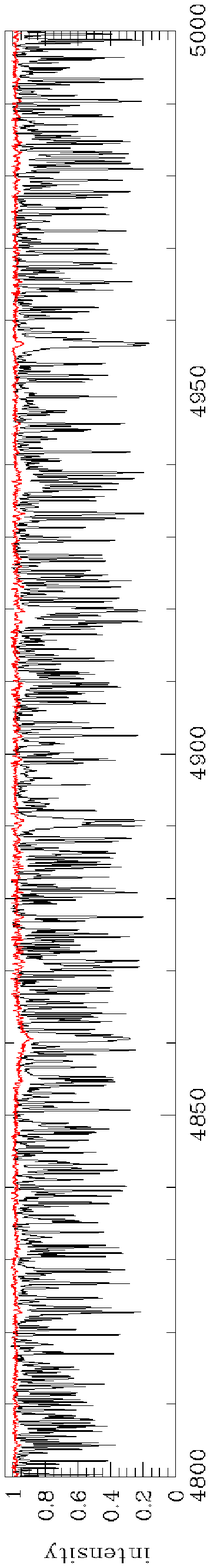}
\includegraphics[angle=-90, width=18.3cm]{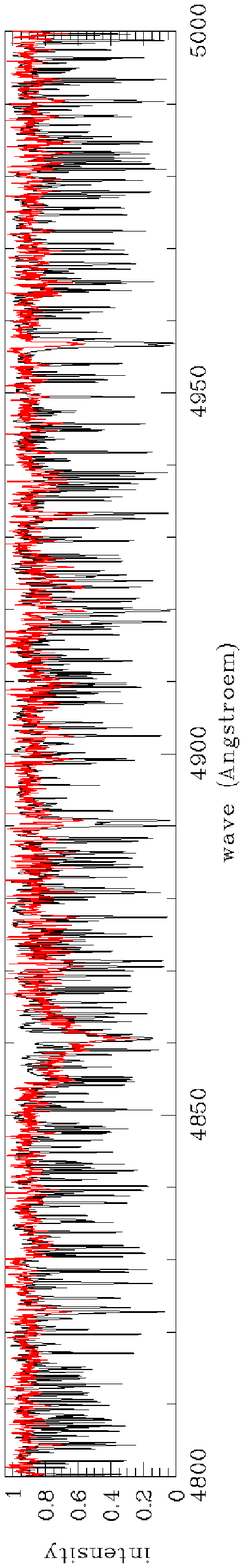}
\caption{Decomposed spectra of the RG (black) and the MS (red) star in the H$\beta$ region.
Top: Normalised to the common continuum of both components. 
Bottom: Normalised to the continua of the single components.}
\label{spectra}
\end{figure*}

Finally, we describe a third method that can be used to extract the orbital parameters for the binary system {\it as well
as disentangled spectra and stellar parameters} for the RG and the MS stars.

\subsection{Using KOREL} \label{disentangle}

We used the {\sc korel} program \citep{korel} to decompose the observed composite spectrum into the spectra of the components, using only the FIES and HERMES spectra because of their higher resolution. {\sc korel} is a Fourier transform-based program for spectral disentangling, i.e.\ it assumes Keplerian motion in a binary or hierarchical multiple system and computes the decomposed spectra together with optimised orbital elements. RVs are computed as well but only have formal values representing the shifts applied to the single spectra to construct the composite spectra. {\sc korel} delivers the decomposed spectra normalised to the common continuum of both components as observed from the composite input spectra. In order to normalise the separated spectra one needs information about the wavelength-dependent flux ratio between the components, which can be obtained from multicolour photometry. We applied Eq.\,1 to perform this normalisation. 

One problem in applying the {\sc korel} program is that the zero frequency mode in the Fourier domain 
is unstable and cannot be determined from the Doppler shifts alone \citep{2008A&A...482.1031H}. 
This effect gives rise to low-frequency undulations in the continua of the decomposed spectra 
and prevents in many cases an accurate determination of the local continuum. The {\sc korel} program
provides the possibility to filter one basic frequency and its harmonics but in practice the
undulations turn out to be too complex. Instead, we applied the program to overlapping
wavelength bins of only 5\,nm each and built the final spectra from these short subframes. Fig.\,\ref{spectra} illustrates, in the upper panel, the faintness of the MS spectrum which leads to the very noisy appearance of the normalised spectrum of this star as shown in the lower panel. The renormalised, decomposed spectra are used for the spectrum analysis. To get orbital parameters comparable to those determined with the other methods, we additionally applied the program to all spectra including the TLS observations. The comparison is given in Table\,\ref{primorbit}.

%
%

\begin{table}
\tabcolsep 1.65mm
\caption{Orbital parameters from the velocities of the RG ($T_0$ is BJD\_TDB 2455000+).}
\begin{tabular}{lllll}
\hline\hline
 & \multicolumn{1}{c}{$T_0$} & \multicolumn{1}{c}{$e$} & \multicolumn{1}{c}{$\omega$} 
 & \multicolumn{1}{c}{$K_{\rm RG}$} \\
 &       &     & \multicolumn{1}{c}{deg.}  & \multicolumn{1}{c}{km~s$^{-1}$} \\
\hline
LBF & 812.19(09) & 0.6888(13) & 300.83(25) & 25.855(69) \\
{\sc todcor}     & 812.19(07) & 0.6887(11) & 300.73(21) & 25.837(56) \\
{\sc korel}      & 812.15(08) & 0.6886(11) & 300.70(27) & 25.808(49) \\
\hline
\end{tabular}
\label{primorbit}
\end{table}

\section{Spectral analysis \label{sect:abund}}

We used the VWA (Versatile Wavelength Analysis) method \citep{vwa} as described in \citet{thygesen} on the disentangled spectrum of the RG star to derive the classical atmospheric parameters \teff\ and \logg\ as well as the metal abundance \feh. The results are presented in Table\,\ref{basics}. From the light curve analysis (Sect.\,\ref{sect:lca}) we determined a more accurate value for \logg\ of 2.57. Plugging this back in the VWA reduction produces an improved value for \feh. Our best estimate for \feh\ is that based on \fetwo, as \feone\ can be affected by non local thermodynamic equilibrium (NLTE) effects in RGs. We note however that the results of \cite{thygesen12} suggest that the results for \feone\ and \fetwo\ should agree for RG stars with parameters similar to the RG in \kic\ when \logg\ is given by an accurate independent measurement. The fact that our \feh$_{\feone}$ and \feh$_{\fetwo}$ values only marginally agree for the RG star could therefore be an indication that the disentangling procedure has introduced small systematic errors in the spectrum. Since this system has a long total eclipse where the MS star is completely occulted by the RG, one could obtain a clean high S/N spectrum of only the RG in order to test such a hypothesis and improve the atmospheric parameters. Knowing the true spectrum of the RG star could also improve RV measurements from present and future spectra. The only issue with obtaining such a spectrum is that it can only be obtained during a short time interval every 408 days during totality of the primary eclipse.

\begin{table}
\caption{Atmospheric parameters from VWA analysis of the RG star.}
\begin{tabular}{ll}
\hline\hline
\teff & $4800\pm80$~K \\
\logg & $2.80\pm0.20$~dex \\
\feh$_{\rm Fe\,{\sc I}}$ & $0.24\pm0.15$~dex (84 lines) \\
\feh$_{\rm Fe\,{\sc II}}$ & $0.21\pm0.15$~dex (8 lines) \\
$v_{\rm micro}$ & $1.35\pm0.15$ \kms \\
$v_{\rm macro}$ & $3.2$ \kms \\
\vsini & $3.2$ \kms \\
\hline
\multicolumn{2}{l}{Using \logg\ from the dEB analysis} \\
\logg & $2.57$ (fixed) \\
\teff\ & not affected \\
\feh$_{\rm Fe\,{\sc I}}$ & $0.20\pm0.11$ (rms)\\
\feh$_{\rm Fe\,{\sc II}}$ & $0.08\pm0.13$ (rms)\\
\hline
\end{tabular}
\label{basics}
\end{table}

The application of VWA presumes the presence of single, unblended lines in the spectrum. Due to the larger $v{\rm sin}i$ and low S/N, it cannot be applied to the decomposed spectrum of the MS star. 
We used the spectrum synthesis method instead, based on \citet{lehmann}. 
This method compares the observed spectrum with synthetic spectra computed on a grid of atmospheric parameters and uses
the $\chi^2$ following from the O-C residual spectra as the measure of the goodness of fit.
The analysis showed that the noisy spectrum
of the MS star only permits the determination of two stellar parameters reliably, namely \teff\ and \vsini. The correlation between all the other parameters listed in Table\,\ref{basics} is very high for the MS star leading to large uncertainties. For that reason we fixed the \logg\ of the MS star to the value of 4.17 as determined from the light curve analysis (Table\,\ref{tab:final} in  Sect.\,\ref{sect:disc}). The $v_{\rm micro}$ is derived for MS stars from \citet{bruntt2} and the $v_{\rm macro}$ from \citet{bruntt3}. Finally, we fixed \feh\ to the RG star's value of 0.1 as derived from its \fetwo\ lines and obtained the results as listed in Table\,\ref{parsec}. In other words we assume that the stars were formed with the same metallicity.

\begin{table}\centering
\caption{Atmospheric parameters for the MS star.}
\begin{tabular}{ll}
\hline\hline
\logg       & ~~~4.17 (taken from light curve analysis)\\
\feh\ (dex) & ~~~0.1 (taken from RG analysis)\\ 
\teff\ ~~~~~(K) &  $6490\pm160$  \\
$v_{\rm micro}$ & $1.6$ \kms\ \citep{bruntt2} \\
$v_{\rm macro}$ & $4.9$ \kms\ \citep{bruntt3} \\
\vsini\ ~~(\kms) & ~$22.2\pm2.5$ \\
\hline
\end{tabular}
\label{parsec}
\end{table}


%

\begin{table}
\centering
\caption{\label{tab:rvfit} Spectroscopic orbital solution for \kic.}
\begin{tabular}{l r@{\,$\pm$\,}l}
\hline
Parameter: & \multicolumn{2}{c}{{\sc sbop} solution} \\
\hline
Measured quantities: \\
$K_{\rm RG}$ (\kms) & 25.85 & 0.07 \\
$K_{\rm MS}$ (\kms) & 30.17 & 0.39 \\
$\gamma_{\rm RG}$ (\kms) & $-46.43$ & 0.04 \\
$\gamma_{\rm MS}$ (\kms) & $-45.66$ & 0.27 \\
$T_0$ (BJD\_TDB\,2455000+) & 812.19 & 0.08 \\
$e$ & 0.6888 & 0.0013 \\
$\omega$ ($ ^{\circ}$) & 300.88 & 0.23 \\
Derived quantities: \\
$M_{\rm RG}\sin^3i$ (\msun) & 1.526 & 0.040 \\
$M_{\rm MS}\sin^3i$ (\msun) & 1.308 & 0.018 \\
$q = M_{\rm MS}/M_{\rm RG}$ & 0.857 & 0.011 \\
$a_{\rm RG}\sin i $ (\rsun) & 176.4 & 2.3 \\
$a_{\rm MS}\sin i $ (\rsun) & 151.2 & 0.1 \\
\hline
\end{tabular}
\label{rv_results}
\end{table}

\section{Orbital analysis \label{sect:orbit}}

The problems raised by the determination of the RVs of the MS star decrease when taking into account that not all orbital phases are needed for this star. The orbit of the RG is so well determined that only two parameters need to be constrained: the velocity semi-amplitude $K_{\rm MS}$ and the systemic velocity $\gamma_{\rm MS}$. Therefore one can easily ignore or at least give low weights to the RVs at phases where it is difficult or impossible to get consistent results from the LBF or {\sc todcor} reductions. Thus, low weights are applied at phases with small RV differences. We decided to let $\gamma_{\rm MS}$ be a free parameter even if \citet{pasquini} find no difference in the RVs between dwarfs and RGs in the open cluster M\,67. Potential differences can come from gravitational redshifts, line asymmetries, and from systematic errors in the extraction of RVs from composite spectra. Fixing the systemic velocity of the MS star to that of the RG star changes the $K_{\rm MS}$ velocity by 0.2\,\kms, leading to a change in mass of the order of 0.015\,\msun. This is included in the final uncertainty for the masses as a systematic uncertainty.

\begin{figure}
\includegraphics[width=10.4cm]{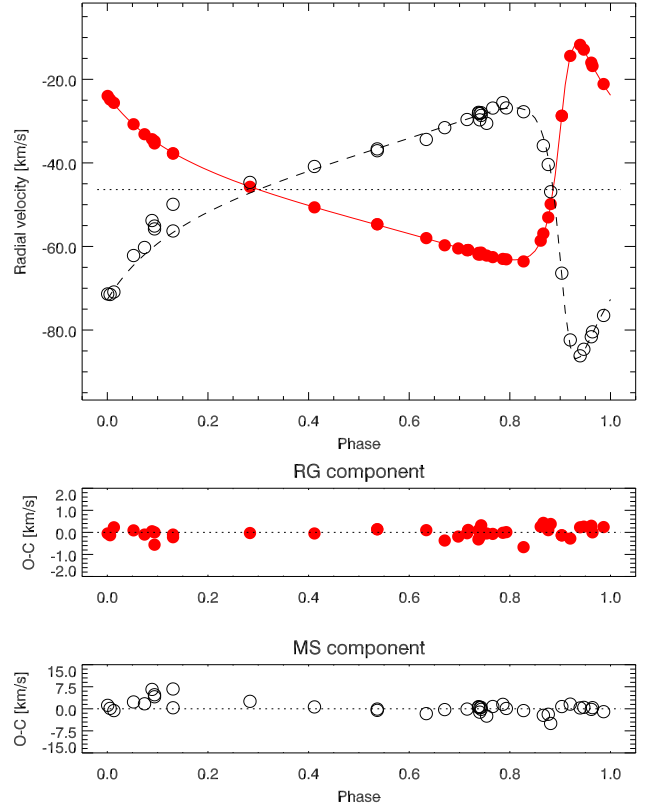}
\caption{Plot of the {\sc sbop} solutions for the two components of \kic. Red dots are for the 
RG, open black dots for the MS star. Note the difference in vertical scale for the O-C plots.}
\label{rvfig_final}
\end{figure}

The solution for the orbit was calculated using the {\sc sbop} code \citep{sbop}. The program needs the RVs and some estimates for the orbital parameters (see Table\,\ref{tab:rvfit}) as input. One can also keep some of the orbital parameters fixed and solve for one or two stellar components or apply weights to the data points.

From the start we froze the value of the orbital period $P$ to that obtained to high precision from the \kepler\ light curve. Then we solved for the RG component to get the common orbital parameters. We used weights based on the statistical uncertainty on each data point. This led to accurate values for the following parameters: the time of periastron $T_0$, the eccentricity $e$, the longitude of periastron $\omega$, and the velocity component $K_{\rm RG}$. It also provided us with the systemic velocity $\gamma_{\rm RG}$ for the orbit of the RG star. As seen from Table~\ref{primorbit}, there is hardly any difference between the LBF and the {\sc todcor} results. The orbital parameters from all three methods used agree very well within the measurement uncertainties. We also solved for both components simultaneously, but the common orbital parameters did not change due to much larger uncertainties on the RVs of the dwarf star.

\begin{figure*} \includegraphics[width=\textwidth,angle=0]{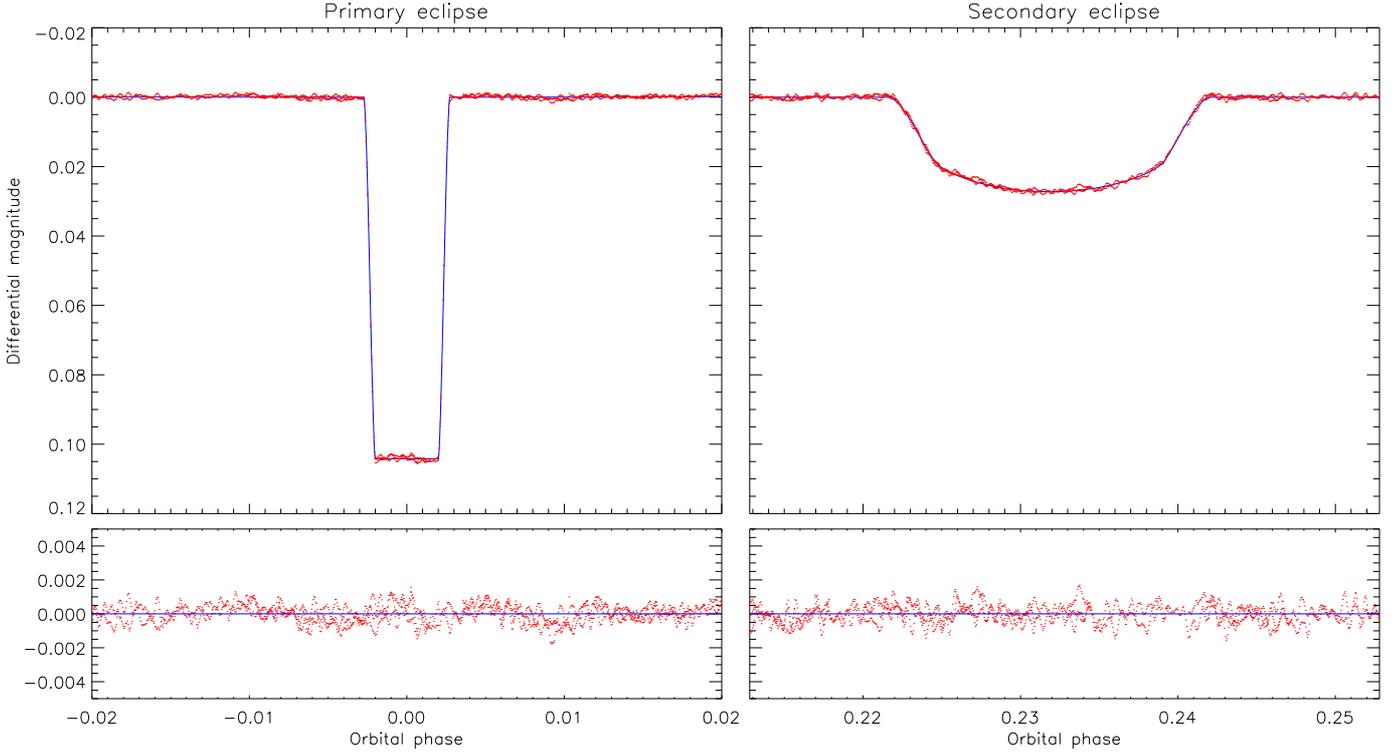} \\
\caption{\label{fig:lcfit} The primary (left panels) and secondary eclipse 
(right panels) of \kic. The {\it Kepler} data points are shown as red dots 
and the {\sc jktebop} best fit as the blue curve. The residuals of the fit 
are shown in the lower panels with an expanded scale on the ordinate axis.} 
\end{figure*}

In the next round we kept the four common parameters fixed and solved only for the orbit of the MS star to get $K_{\rm MS}$ and $\gamma_{\rm MS}$. The weights applied here are the weights derived in Sect.\,\ref{sect:rv}. They were based on the robustness of the RV determination and emphasise the data points with large RV separations between the RG and the MS star, giving  high weights to the extremes. Focusing on parts of the orbit is justified by the fact that we have 43 RVs and only need to determine two parameters, so we are able to reject or down-weight a significant fraction of the data points. The TLS data, in particular, have less coverage in the blue part of the spectrum than the {\textsc Mercator} or NOT data. They also contain a few cases of obvious outliers at epochs where we did expect problems. We discarded these observations.

The final results of the analysis of the spectra are listed in Table\,\ref{rv_results}, where we took the average of the LBF and the {\sc todcor} results from Table\,\ref{primorbit}. {\sc korel} does not directly give velocities but only the orbital parameters and was considered less precise. The fit of the RVs\footnote{The RVs and the weights applied can be obtained from the first author} is shown in Fig.\,\ref{rvfig_final}. The mean scatter around the solution is 0.2\,\kms\ per data point for the RG star and 1.1\,\kms\ for the MS star. The uncertainties on the stellar masses come mainly from $K_{\rm MS}$, as the uncertainty on $K_{\rm MS}$ is five times the uncertainty on $K_{\rm RG}$. To account for the precision of the systematic errors we add a systematic uncertainty of 0.2\kms\ as explained earlier in this section. This additional uncertainty is included in Table\,\ref{rv_results}. The {\sc sbop} fit for the RG star is very good as shown by the residuals, whereas for the MS star there are clearly problems at certain phases.

As a last check of the results presented in Table\,\ref{rv_results}, a bootstrap method \citep{bootstrap} based
on a least square solution for all seven parameters simultaneously was applied to the same radial velocities using
the same weights. The results of this give values that have a mean difference of $0.3\sigma$ and a mean ratio
of $\sigma$'s of 1.17. This difference is considered acceptable as it will not affect the conclusions.

\section{Light curve analysis} \label{sect:lca}

The \kepler\ light curve of \kic\ shows variability due to oscillations and eclipses. As discussed at the end of Sect.\,\ref{sec:intro}, we adopt the convention that the primary eclipse is deeper than the secondary. The primary eclipse is roughly 0.105\,mag deep, and occurs when the MS star passes behind the RG. As the radius of the RG is much greater than that of the MS star, the primary eclipse is total and has a flat base. The secondary eclipse is about 0.028\,mag deep, and is caused by the passage of the MS star in front of the RG. It is an annular eclipse (a transit), and shows a curved base due to limb darkening.

The available {\it Kepler} data cover three primary eclipses (Q1, Q6, Q10) and three secondary eclipses (Q2, Q7, Q11). One of the primary eclipses and two of the secondary eclipses have incomplete coverage due to spacecraft safe-mode events and other phenomena. Before performing the light curve analysis we needed to extract the eclipses and normalise each of them to unit flux, as the absolute fluxes recorded by {\it Kepler} are subject to instrumental variations on monthly timescales. Each eclipse was therefore extracted from the full light curve, accompanied by the data taken immediately before and after. We retained those data points which were less than roughly one eclipse duration before the start or after the end of eclipse. The out-of-eclipse data were then fitted with low-order polynomials in order to normalise each eclipse to unit flux. We tried using different polynomial orders, from first to third, and found that this has an effect of at most $0.2\,\sigma$ on our resulting photometric parameters.

Once the regions around each eclipse had been collected and normalised to unit flux, we converted the data into magnitudes. This light curve was then modelled using the {\sc jktebop}\footnote{{\sc jktebop} is written in {\sc fortran77} and the source code is available at {\tt http://www.astro.keele.ac.uk/jkt/codes/jktebop.html}} code \citep{jktebop} to obtain the photometric parameters. {\sc jktebop} considers the component stars to be biaxial spheroids and is well suited to a system like \kic\ which contains almost-spherical stars. The model was numerically integrated over time intervals of 1765\,s to match the duration of the {\it Kepler} long-cadence data \citep{john}.

We fitted for the orbital period, $P_{\rm orb}$, the midpoint of the first primary eclipse, $T_{ecl}$, the orbital inclination, $i$, and the fractional radii of the two stars, $r_{\rm MS}$ and $r_{\rm RG}$. The fractional radii are defined to be the true radii divided by the orbital semi-major axis, and are quantities which can be obtained directly from the light curve. We followed the {\sc jktebop} default of fitting for the sum and ratio of the fractional radii. We also fitted for $e$ and $\omega$, via the combination terms $e\cos\omega$ and $e\sin\omega$. We ran a test fit including contaminating `third' light and found that this parameter was very close to zero. We therefore fixed it at zero for the final fit. 

Limb darkening (LD) was accounted for using the quadratic law, after tests with the linear law returned a marginally poorer fit. We fitted for the linear LD coefficient for each star, $u_{\rm MS}$ and $u_{\rm RG}$. The nonlinear LD coefficients were fixed at theoretically predicted values of $v_{\rm MS} = 0.14$ and $v_{\rm RG} = 0.32$ \citep{Sing10aa} as they are strongly correlated with the linear coefficients \citep{Me++07aa}. Alternative fits using other two-coefficient LD laws returned results in very close agreement with those from the quadratic law.

Our initial fits returned values of $e\cos\omega$ and $e\sin\omega$ which were inconsistent with those found from the RVs in the preceding sections. This is unsurprising because the shape of the spectroscopic orbit is much more sensitive than the light curve to the value of $e\sin\omega$. We therefore modified {\sc jktebop} to simultaneously fit for the {\it Kepler} light curve and our measured RVs. The resulting best fit is closer to the spectroscopic than the photometric results, and is a good fit to all data. We find rms residuals of 0.52\,mmag for the light curve, 0.22\,\kms\ for the RVs of the giant and 0.97\,\kms\ for the RVs of the MS star.

The best fit is shown in Fig.\,\ref{fig:lcfit} which gives also a good representation of the eclipse shapes. The residuals of the fit, however, show small trends with timescales of hours and days. The former are due to oscillations in the giant star, which are not included in the {\sc jktebop} model. The latter could be $\gamma$\,Doradus pulsations in the MS star, whose \teff\ is close to the red edge of the $\gamma$\,Doradus instability strip. The presence of these effects means that classical error analysis techniques, which rely on data points being independent and identically distributed, will likely underestimate the uncertainties on the measured parameters. 

We therefore obtained parameter uncertainties using a residual-permutation algorithm, as implemented by \citet{exolc}. Those for the photometric parameters are 2--3 times larger than the uncertainties from a Monte Carlo analysis (see \citealt{wwaur}), confirming our suspicions above. We then inspected plots of the light curves generated for a range of values for each fitted parameter. As a general rule the fit deteriorates noticeably by the time a parameter deviates by twice the uncertainty given by the residual-permutation algorithm. We therefore conservatively adopt twice the residual-permutation error bars as the overall uncertainties in the parameter measurements. These numbers should be interpreted as $1\sigma$ uncertainties. Further observations in the near future by the {\it Kepler} satellite will allow these uncertainties to be pushed down. The spectroscopic parameters ($K_{\rm MS}$, $K_{\rm RG}$, $\gamma_{\rm MS}$ and $\gamma_{\rm RG}$) did not show significantly larger uncertainties from the residual-permutation algorithm than from the Monte Carlo algorithm. We therefor did not double their uncertainties, but instead adopted the larger of the values from the Monte Carlo and residual-permutation algorithms.

\begin{table} \centering
\caption{\label{tab:lcfit} Parameters obtained from the light curve modelling of \kic\ with {\sc jktebop}.
Units indicated if parameter is not dimensionless.}
\begin{tabular}{l r@{\,$\pm$\,}l} \hline
Parameter                     & \multicolumn{2}{c}{Value} \\
\hline
$P_{\rm orb}$ (d)             & 408.3241  & 0.0010    \\
$T_{ecl}-2400000$ (BJD\_TDB)  & 54990.6197& 0.0014    \\
$i$ (degrees)                 & 89.602    & 0.078     \\
$r_{\rm MS}+r_{\rm RG}$       & 0.037341  & 0.00037   \\
$r_{\rm RG}/ r_{\rm MS}$      & 6.833     & 0.073     \\
$e\cos\omega$                 & $-$0.3517 & 0.0017    \\
$e\sin\omega$                 & 0.5895    & 0.0054    \\
$u_{\rm MS}$                  & 0.07      & 0.11      \\
$u_{\rm RG}$                  & 0.533     & 0.032     \\
$K_{\rm MS}$ (\kms)           & 30.36     & 0.24      \\
$K_{\rm RG}$ (\kms)           & 25.782    & 0.085     \\
$\gamma_{\rm MS}$ (\kms)      & $-$45.43  & 0.20      \\
$\gamma_{\rm RG}$ (\kms)      & $-$46.44  & 0.03      \\[3pt]
$r_{\rm MS}$                  & 0.004767  & 0.00009   \\
$r_{\rm RG}$                  & 0.032574  & 0.00030   \\
$e$                           & 0.6864    & 0.0019    \\
$\omega$ (degrees)            & 300.82    & 0.35      \\          
$\ell_{\rm RG}/\ell_{\rm MS}$ & 9.927     & 0.028     \\
\hline \end{tabular} 
\end{table}

The final parameters and uncertainties are given in Table\,\ref{tab:lcfit}. The upper quantities in the table were fitted by {\sc jktebop} and the lower five quantities are derived parameters. The light ratio, $\ell_{\rm RG}/\ell_{\rm MS}$, is for the \kepler\ passband. Our definition that the primary eclipse is deeper than the secondary eclipse means that the ratio of the radii is greater than unity. The $T_{ecl}$ has been corrected for the timing error in the \kepler\ timestamps\footnote{See http://archive.stsci.edu/kepler/timing\_error.html}. The timings supplied with the {\it Kepler} data are corrected to the barycentre of the Solar system and are on the TDB timescale\footnote{See the 2012 March edition of the {\it Kepler} Data Charac\-teristics Hand\-book at {\tt http://archive.stsci.edu/kepler/} {\tt manuals/Data\_Characteristics.pdf}.}.

\section{Physical properties and distance test} \label{sect:dist}

\begin{table}
\centering
\caption{Physical parameters of the components.}
\begin{tabular}{l r@{\,$\pm$\,}l}
\hline
Parameter: & \multicolumn{2}{c}{value} \\
\hline
a (\rsun) & 329.6 & 1.7 \\
$M_{\rm RG}$ (\msun) & 1.557 & 0.028 \\
$M_{\rm MS}$ (\msun) & 1.322 & 0.017 \\
$R_{\rm RG}$ (\rsun) & 10.74 & 0.11 \\
$R_{\rm MS}$ (\rsun) & 1.571 & 0.031 \\
\teff$_{\rm ,RG}$ (K) & 4800 & 80 \\
\teff$_{\rm ,MS}$ (K) & 6490 & 160 \\
\logg$_{\rm RG}$ & 2.569 & 0.009 \\
\logg$_{\rm MS}$ & 4.167 & 0.016 \\
$\log L_{\rm RG}$/\lsun & 1.739 & 0.030 \\
$\log L_{\rm MS}$/\lsun & 0.610 & 0.031 \\ 
$E(B-V)$ (mag) & 0.14 & 0.05 \\
Distance (pc) & 998 & 21 \\
\hline
\end{tabular}
\label{tab:final}
\end{table}

We have calculated the physical properties and distance of \kic\ using the {\sc absdim} code \citep{southworth}. This code calculates physical properties using standard formulae, and distance measurements for the system using several methods. The results of this process are given in Table\,\ref{tab:final} and show that the masses and radii of the two stars are measured to precisions of 2\% or better. We adopted the values of $r_{\rm MS}, r_{\rm RG}$, $e$, $K_{\rm MS}$ and $K_{\rm RG}$ from Table\,\ref{tab:lcfit}. For the \teff\ of the RG we adopt that determined from its disentangled spectrum (Sect.\,\ref{sect:abund}). For the MS star's \teff\ we take the mean of the values from disentangling and from photometry with an uncertainty sufficient to encompass both measurements.

The two components of \kic\ must be at the same distance from Earth. This provides a consistency check on the \teff s and the radii of the two objects. In this case we use it to check the \teff\ of the MS star, as the estimates derived above are quite uncertain. The $B$-band apparent magnitude of the system was taken from the Tycho catalogue \citep{Hog+97aa}, and $JHK_s$ magnitudes from 2MASS \citep{Cutri+03book}. The Tycho $V$ magnitude is rather faint at $11.33 \pm 0.08$, and cannot be reconciled simultaneously with $JHK$ and $B$ whatever the distance and reddening. The $g$ and $r$ band photometry in the Kepler Input Catalogue \citep{Brown+11aj} and the calibration from \citet{Jester+05aj} suggest a value of $V = 11.06 \pm 0.10$, where the uncertainty is a conservative guess. We have used this as a consistency check but not to calculate any results presented in the current work. We do not have apparent magnitudes in the $R$ or $I$ bands, which would be useful if they existed. 

The distance to the system was calculated in each passband, and the reddening $E(B-V)$ was adjusted until the optical and infrared distances were aligned. We obtained $E(B-V) = 0.14 \pm 0.05$ where the uncertainty is conservative. This value depends primarily on the $B$-band apparent magnitude. The distances from the surface brightness method (see \citealt{southworth}) and from the usual bolometric correction method are in acceptable agreement, where we took bolometric corrections from \citet{bessell} for $B$ and $K$, or \citet{girardi} for $UBVRIJHK$. The best distance is determined from the $K$ band, due to the more precise apparent magnitudes and the lesser influence of reddening, and is $d = 997 \pm 21$\,pc. 

Then we added the $BVRI$ light ratios, obtained from the fits to the $BVRI$ light curves. This allowed us to calculate distances to each component separately, for those passbands for which we possess both an apparent magnitude and a flux ratio ($B$ and $V$). With the \teff s given above, we found slightly discrepant distances for the two stars which could be nullified by adopting \teff$_{\rm ,MS} = 6670$\,K for the MS star. This has very little effect on the distance we find ($+1$\,pc in the $K$ band), which depends mostly on the RG due to its much larger size. Our final distance is $998 \pm 23$\,pc, from the surface brightness method in the $K$ band. For comparison, the bolometric-correction based distances are a little larger at $1015 \pm 20$\,pc for the \citet{bessell} and $1029 \pm 20$\,pc for the \citet{girardi} tabulations.

\section{Discussion \label{sect:disc}}

\begin{figure} \label{rm_plot}
\includegraphics[width=9.5cm,angle=0]{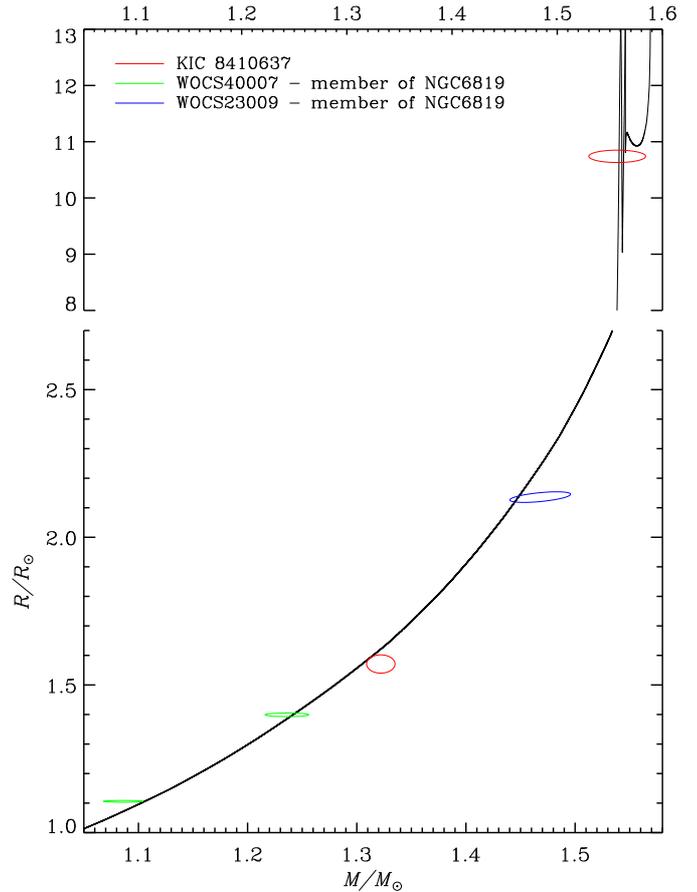} 
\caption{\label{fig:rm} Results for three dEB systems: \kic\ (red ellipses), and 
WOCS40007 and WOCS23009 in NGC\,6819 (green and blue ellipses). Ellipses indicate 
$1\,\sigma$ uncertainties. The isochrone is for an age of $t = 2.60$\,Gy, see text.}
\end{figure}



\begin{figure} 
\includegraphics[width=9.0cm,angle=0]{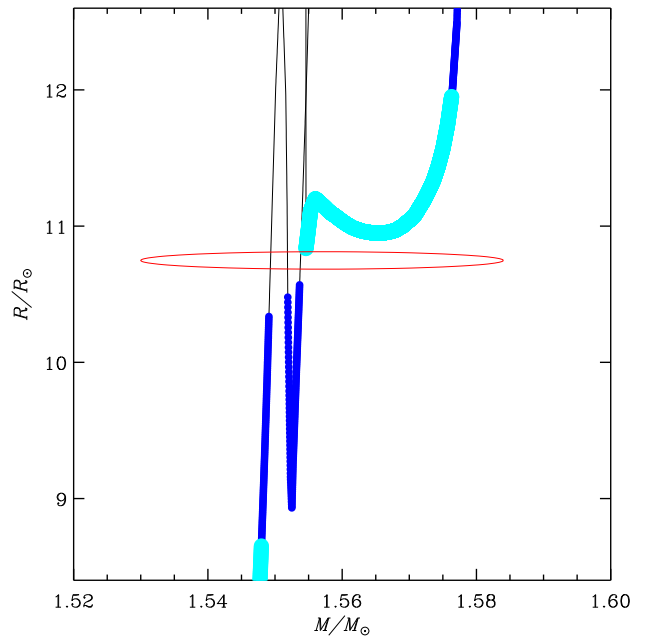} 
\caption{\label{fig:rm2} 
Zoom on the upper part of Fig.\,\ref{fig:rm}. The isochrone is coloured 
cyan or blue in regions where the isochrone \teff\ is within 0--1$\sigma$ 
or 1--2$\sigma$, respectively, of the measured \teff\ of the RG.}
\end{figure}

The position of the two objects in the mass-radius plane is shown in Fig.\,\ref{fig:rm}. The confusion of isochrone lines going through the measured properties of the RG component is due to the fact that its position is consistent with several evolutionary stages: just before, during, or just after the bump phase on the RGB or the RC core helium-burning phase after the helium-flash. We have therefore zoomed in on this part of the diagram in Fig.\,\ref{fig:rm2}, and have coloured the isochrone around the RG measurement. This strongly suggests that the RG is in the RC phase, since this is the only evolutionary phase with a consistent \teff\ within the uncertainty ellipse, even considering 2$\sigma$ uncertainties. The lower coloured parts, which correspond to the RGB phase, does not quite reach the uncertainty ellipse.


Evolutionary timescales provide additional support for this scenario, as the RC phase is much longer than the bump phase. Furthermore, by assuming that the star is an RC star, and using calibrations of the absolute infrared magnitudes of RC stars (e.g.\ \citealt{Salaris02}) we obtained a distance for \kic\ in excellent agreement with that of Sect.\,7.
The individual oscillation frequencies derived from the latest combined \kepler\ data sets are needed to get asteroseismic parameters leading to strict constraints on the stellar evolution. At present, the power spectrum of \kic\ does not allow the evolutionary status of the RG to be pinned down to the RGB or RC phases.
If the conclusion that the RG star is a RC star is confirmed, we have to address the following problem:
the MS and the RG stars have a periastron distance of 103\rsun (see Table\,\ref{basics}). This is comparable to the size of the RG at the tip of the giant branch in Fig.\,\ref{fig:rm}). Will significant mass transfer take place at that period modifing the original masses of the MS and RG stars? 

\kic\ seems to have reached the same evolutionary state as the open cluster NGC\,6819 as shown by Fig.\,\ref{fig:rm}.
Fig.\,\ref{fig:rm} includes three stars from dEB systems in NGC\,6819 (\citealt{sandquist}, Jefferies et al.\ in prep). NGC\,6819 is located in the \kepler\ field and has $\feh = 0.09$ \citep{brag2}, similar to \kic. The isochrone shown is a {\sc parsec} model for $\feh = 0.08$ and $t = 2.60$\,Gy \citep{parsec}. Since {\it both components} of \kic\ match the mass-radius relation of NGC\,6819, indications are that \kic\ has an age very similar to NGC\,6819. The similarity of the values of the asteroseismic large separation, \logg, \teff, 2MASS $J-K_s$ colour, and reddening of \kic\ to those of RC stars in NGC\,6819 (\citealt{Corsaro12,brag2,Cutri+03book,sandquist}) provides further support to our finding that that the RG in \kic\ is an RC star.


Stars on the RGB (and also the RC) in NGC\,6819 have masses in the range 1.52--1.56\msun\ according to \citet{sandquist}, in agreement with the mass of the RG star in \kic. 
The peak in the mass distribution for NGC\,6819 in Fig.\,2 in \citet{clusters} is also located at a mass close to 1.5\msun.
This is significantly less than the mass $1.68\pm0.03$\msun\ derived for RG stars in NGC\,6819 by \citet{ngc6819} using asteroseismic grid modelling with \kepler\ light curves. \citet{sandquist} have no direct determination of the mass of a RG star, but \kic\ supports their mass estimate as the masses and radii of both components fit nicely on an isochrone for the cluster. A corresponding difference between the masses of RG stars derived from observations of dEBs and asteroseismic parameters was reported by \citet{ngc6791_2} for the old metal-rich open cluster NGC\,6791. Interestingly, these discrepancies disappear if one uses the most up-to-date version of the scaling relations \citep{mosser13} to derive the asteroseismic masses of the cluster stars. 

It will be very interesting to compare the power spectra of the \kepler\ light curves for \kic\ and similar RG stars in NGC\,6819 to investigate possible similarities.
This is however beyond the scope of the current work.

For now, comparing the \logg\ values of the RG star derived from asteroseismic scaling relations ($2.6\pm0.1$; \citealt{hekker}) and the present binary analysis ($2.569\pm0.009$) provides the first indication that surface gravities for RGs from the asteroseismic scaling relations are not only precise but also accurate. 

\section{Conclusion \label{sect:concl}}

The main points are summarised here.
\begin{itemize}
\item
\kic\ is the first of a set of very long period eclipsing binaries with \kepler\ light curves to be subject to detailed study. From \kepler\ and ground-based photometry combined with high-resolution spectroscopy, accurate masses and radii of the component stars have been derived using standard tools for measuring radial velocities and for analysing light curves of dEBs.
\item
The system has a high eccentricity ($e \sim 0.7$) which, despite the long orbital period of 408 days, leads to a small orbital separation at periastron. We find that the RG has passed through the RGB phase and is now an RC star. The two components would have come into contact when the RG was near the RGB tip.
\item
The two components of \kic\ fit the mass-radius relation of the open cluster NGC\,6819 (Fig.\,\ref{fig:rm}), and have a metallicity very similar to that of the cluster. This strongly suggests that the age of \kic\ is very similar to that of NGC\,6819. The power spectrum of the \kepler\ data of \kic\ should be very similar to power spectra of RG stars with similar parameters in NGC\,6819, making a future comparative asteroseismic analysis interesting. 
\item
There are many more dEBs with RG stars among the \kepler\ targets, which can be used to determine accurate masses and radii for more RGs. A suitable system, which has a shorter period of $P=68$\,d and thus is easier to observe as one can collect the necessary data in one semester, is KIC\,10015516 from \citet{ebcat2}.
\end{itemize}
We are just at the beginning of using dEBs to provide basic calibrations of some of the many RG stars observed by \kepler. The prospects for understanding the structure of evolved stars and obtaining more accurate ages for stars and open clusters look very promising.

\begin{acknowledgements}
This research took advantage of the {\it Simbad} and {\it Vizier} databases at the CDS, Strasbourg (France), and NASA's Astrophysics Data System Bibliographic Services.
JS acknowledges financial support from STFC in the form of an Advanced Fellowship.
Based on observations obtained with the HERMES spectrograph, which is supported by the Fund for Scientific Research of Flanders (FWO), Belgium, the Research Council of K.U.Leuven, Belgium, the Fonds National Recherches Scientific (FNRS), Belgium, the Royal Observatory of Belgium, the Observatoire de Gen\'eve, Switzerland and the Th\"uringer Landessternwarte Tautenburg, Germany.
AP, GK and ZK acknowledge the support from the NCN grant No.~2011/03/B/ST9/02667.
AOT acknowledges support from Sonderforschungsbereich SFB 881 "The Milky Way System" (subproject A5) of the German Research Foundation (DFG). 
KB acknowledges support from the Carlsberg Foundation and the Villum Foundation.
Funding for the Stellar Astrophysics Centre is provided by The Danish National Research Foundation. The research is supported by the ASTERISK project (ASTERoseismic Investigations with SONG and Kepler) funded by the European Research Council (Grant agreement no.\,267864).
SH acknowledge financial support from the Netherlands Organization of Scientific Research (NLO).
PB's research leading to these results has received funding from the
European Research Council under the European Community's Seventh
Framework Programme (FP7/2007--2013)/ERC grant agreement
n$^\circ$227224 (PROSPERITY). JD received funding from the Belgian federal science policy office
(C90309: CoRoT Data Exploitation)
\end{acknowledgements}

\bibliographystyle{aa}
\bibliography{giant_eb-v12} 

\begin{thebibliography}{62}
\expandafter\ifx\csname natexlab\endcsname\relax\def\natexlab#1{#1}\fi

\bibitem[{{Andersen} {et~al.}(1991){Andersen}, {Clausen}, {Nordstrom},
  {Tomkin}, \& {Mayor}}]{tzfor}
{Andersen}, J., {Clausen}, J.~V., {Nordstrom}, B., {Tomkin}, J., \& {Mayor}, M.
  1991, \aap, 246, 99

\bibitem[{{Basu} {et~al.}(2011){Basu}, {Grundahl}, {Stello}, {Kallinger},
  {Hekker}, {Mosser}, {Garc{\'{\i}}a}, {Mathur}, {Brogaard}, {Bruntt},
  {Chaplin}, {Gai}, {Elsworth}, {Esch}, {Ballot}, {Bedding}, {Gruberbauer},
  {Huber}, {Miglio}, {Yildiz}, {Kjeldsen}, {Christensen-Dalsgaard},
  {Gilliland}, {Fanelli}, {Ibrahim}, \& {Smith}}]{ngc6819}
{Basu}, S., {Grundahl}, F., {Stello}, D., {et~al.} 2011, \apjl, 729, L10

\bibitem[{{Beck} {et~al.}(2012){Beck}, {Montalban}, {Kallinger}, {De Ridder},
  {Aerts}, {Garc{\'{\i}}a}, {Hekker}, {Dupret}, {Mosser}, {Eggenberger},
  {Stello}, {Elsworth}, {Frandsen}, {Carrier}, {Hillen}, {Gruberbauer},
  {Christensen-Dalsgaard}, {Miglio}, {Valentini}, {Bedding}, {Kjeldsen},
  {Girouard}, {Hall}, \& {Ibrahim}}]{beck}
{Beck}, P.~G., {Montalban}, J., {Kallinger}, T., {et~al.} 2012, \nat, 481, 55

\bibitem[{{Bedding} {et~al.}(2011){Bedding}, {Mosser}, {Huber},
  {Montalb{\'a}n}, {Beck}, {Christensen-Dalsgaard}, {Elsworth},
  {Garc{\'{\i}}a}, {Miglio}, {Stello}, {White}, {De Ridder}, {Hekker}, {Aerts},
  {Barban}, {Belkacem}, {Broomhall}, {Brown}, {Buzasi}, {Carrier}, {Chaplin},
  {di Mauro}, {Dupret}, {Frandsen}, {Gilliland}, {Goupil}, {Jenkins},
  {Kallinger}, {Kawaler}, {Kjeldsen}, {Mathur}, {Noels}, {Aguirre}, \&
  {Ventura}}]{bedding1}
{Bedding}, T.~R., {Mosser}, B., {Huber}, D., {et~al.} 2011, \nat, 471, 608

\bibitem[{{Bessell} {et~al.}(1998){Bessell}, {Castelli}, \& {Plez}}]{bessell}
{Bessell}, M.~S., {Castelli}, F., \& {Plez}, B. 1998, \aap, 333, 231

\bibitem[{{Bragaglia} {et~al.}(2001){Bragaglia}, {Carretta}, {Gratton}, {Tosi},
  {Bonanno}, {Bruno}, {Cal{\`i}}, {Claudi}, {Cosentino}, {Desidera},
  {Farisato}, {Rebeschini}, \& {Scuderi}}]{brag2}
{Bragaglia}, A., {Carretta}, E., {Gratton}, R.~G., {et~al.} 2001, \aj, 121, 327

\bibitem[{{Bressan} {et~al.}(2012){Bressan}, {Marigo}, {Girardi}, {Salasnich},
  {Dal Cero}, {Rubele}, \& {Nanni}}]{parsec}
{Bressan}, A., {Marigo}, P., {Girardi}, L., {et~al.} 2012, \mnras, 427, 127

\bibitem[{{Brogaard} {et~al.}(2011){Brogaard}, {Bruntt}, {Grundahl}, {Clausen},
  {Frandsen}, {Vandenberg}, \& {Bedin}}]{ngc6791_1}
{Brogaard}, K., {Bruntt}, H., {Grundahl}, F., {et~al.} 2011, \aap, 525, A2

\bibitem[{{Brogaard} {et~al.}(2012){Brogaard}, {VandenBerg}, {Bruntt},
  {Grundahl}, {Frandsen}, {Bedin}, {Milone}, {Dotter}, {Feiden}, {Stetson},
  {Sandquist}, {Miglio}, {Stello}, \& {Jessen-Hansen}}]{ngc6791_2}
{Brogaard}, K., {VandenBerg}, D.~A., {Bruntt}, H., {et~al.} 2012, \aap, 543,
  A106

\bibitem[{{Brown} {et~al.}(2011){Brown}, {Latham}, {Everett}, \&
  {Esquerdo}}]{Brown+11aj}
{Brown}, T.~M., {Latham}, D.~W., {Everett}, M.~E., \& {Esquerdo}, G.~A. 2011,
  AJ, 142, 112

\bibitem[{{Bruntt} {et~al.}(2012){Bruntt}, {Basu}, {Smalley}, {Chaplin},
  {Verner}, {Bedding}, {Catala}, {Gazzano}, {Molenda-{\.Z}akowicz}, {Thygesen},
  {Uytterhoeven}, {Hekker}, {Huber}, {Karoff}, {Mathur}, {Mosser},
  {Appourchaux}, {Campante}, {Elsworth}, {Garc{\'{\i}}a}, {Handberg},
  {Metcalfe}, {Quirion}, {R{\'e}gulo}, {Roxburgh}, {Stello},
  {Christensen-Dalsgaard}, {Kawaler}, {Kjeldsen}, {Morris}, {Quintana}, \&
  {Sanderfer}}]{bruntt2}
{Bruntt}, H., {Basu}, S., {Smalley}, B., {et~al.} 2012, \mnras, 423, 122

\bibitem[{{Bruntt} {et~al.}(2010){Bruntt}, {Bedding}, {Quirion}, {Lo Curto},
  {Carrier}, {Smalley}, {Dall}, {Arentoft}, {Bazot}, \& {Butler}}]{bruntt3}
{Bruntt}, H., {Bedding}, T.~R., {Quirion}, P.-O., {et~al.} 2010, \mnras, 405,
  1907

\bibitem[{{Bruntt} {et~al.}(2004){Bruntt}, {Bikmaev}, {Catala}, {Solano},
  {Gillon}, {Magain}, {Van't Veer-Menneret}, {St{\"u}tz}, {Weiss}, {Ballereau},
  {Bouret}, {Charpinet}, {Hua}, {Katz}, {Ligni{\`e}res}, \& {Lueftinger}}]{vwa}
{Bruntt}, H., {Bikmaev}, I.~F., {Catala}, C., {et~al.} 2004, \aap, 425, 683

\bibitem[{{Chernick}(2007)}]{bootstrap}
{Chernick}, M.~R. 2007, {Bootstrap Methods: A Guide for Practitioners and
  Researchers} (Wiley)

\bibitem[{{Clausen} {et~al.}(2010){Clausen}, {Frandsen}, {Bruntt}, {Olsen},
  {Helt}, {Gregersen}, {Juncher}, \& {Krogstrup}}]{bkpeg}
{Clausen}, J.~V., {Frandsen}, S., {Bruntt}, H., {et~al.} 2010, \aap, 516, A42

\bibitem[{{Corsaro} {et~al.}(2012){Corsaro}, {Stello}, {Huber}, {Bedding},
  {Bonanno}, {Brogaard}, {Kallinger}, {Benomar}, {White}, {Mosser}, {Basu},
  {Chaplin}, {Christensen-Dalsgaard}, {Elsworth}, {Garc{\'{\i}}a}, {Hekker},
  {Kjeldsen}, {Mathur}, {Meibom}, {Hall}, {Ibrahim}, \& {Klaus}}]{Corsaro12}
{Corsaro}, E., {Stello}, D., {Huber}, D., {et~al.} 2012, \apj, 757, 190

\bibitem[{{Cutri} {et~al.}(2003){Cutri}, {Skrutskie}, {van Dyk}, {Beichman},
  {Carpenter}, {Chester}, {Cambresy}, {Evans}, {Fowler}, {Gizis}, \&
  {Others}}]{Cutri+03book}
{Cutri}, R.~M., {Skrutskie}, M.~F., {van Dyk}, S., {et~al.} 2003, {2MASS All
  Sky Catalogue of point sources.} (The IRSA 2MASS All-Sky Point Source
  Catalogue, NASA/IPAC Infrared Science Archive)

\bibitem[{{Derekas} {et~al.}(2007){Derekas}, {Kiss}, \& {Bedding}}]{macho}
{Derekas}, A., {Kiss}, L.~L., \& {Bedding}, T.~R. 2007, \apj, 663, 249

\bibitem[{{Devor}(2005)}]{ogle}
{Devor}, J. 2005, \apj, 628, 411

\bibitem[{{Devor} {et~al.}(2008){Devor}, {Charbonneau}, {O'Donovan},
  {Mandushev}, \& {Torres}}]{trescat}
{Devor}, J., {Charbonneau}, D., {O'Donovan}, F.~T., {Mandushev}, G., \&
  {Torres}, G. 2008, \aj, 135, 850

\bibitem[{{Eastman} {et~al.}(2010){Eastman}, {Siverd}, \& {Gaudi}}]{bjd_tdb}
{Eastman}, J., {Siverd}, R., \& {Gaudi}, B.~S. 2010, \pasp, 122, 935

\bibitem[{{Etzel}(1981)}]{sbop}
{Etzel}, P.~B. 1981, in Photometric and Spectroscopic Binary Systems, ed. E.~B.
  {Carling} \& Z.~{Kopal}, 111

\bibitem[{{Gaulme} {et~al.}(2013){Gaulme}, {McKeever}, {Rawls}, {Jackiewicz},
  {Mosser}, \& {Guzik}}]{gaulme}
{Gaulme}, P., {McKeever}, J., {Rawls}, M.~L., {et~al.} 2013, \apj, 767, 82

\bibitem[{{Girardi} {et~al.}(2002){Girardi}, {Bertelli}, {Bressan}, {Chiosi},
  {Groenewegen}, {Marigo}, {Salasnich}, \& {Weiss}}]{girardi}
{Girardi}, L., {Bertelli}, G., {Bressan}, A., {et~al.} 2002, \aap, 391, 195

\bibitem[{{Hadrava}(1997)}]{korel}
{Hadrava}, P. 1997, \aaps, 122, 581

\bibitem[{{Hekker} {et~al.}(2011{\natexlab{a}}){Hekker}, {Basu}, {Stello},
  {Kallinger}, {Grundahl}, {Mathur}, {Garc{\'{\i}}a}, {Mosser}, {Huber},
  {Bedding}, {Szab{\'o}}, {De Ridder}, {Chaplin}, {Elsworth}, {Hale},
  {Christensen-Dalsgaard}, {Gilliland}, {Still}, {McCauliff}, \&
  {Quintana}}]{clusters}
{Hekker}, S., {Basu}, S., {Stello}, D., {et~al.} 2011{\natexlab{a}}, \aap, 530,
  A100

\bibitem[{{Hekker} {et~al.}(2010){Hekker}, {Debosscher}, {Huber}, {Hidas}, {De
  Ridder}, {Aerts}, {Stello}, {Bedding}, {Gilliland}, {Christensen-Dalsgaard},
  {Brown}, {Kjeldsen}, {Borucki}, {Koch}, {Jenkins}, {Van Winckel}, {Beck},
  {Blomme}, {Southworth}, {Pigulski}, {Chaplin}, {Elsworth}, {Stevens},
  {Dreizler}, {Kurtz}, {Maceroni}, {Cardini}, {Derekas}, \& {Suran}}]{hekker}
{Hekker}, S., {Debosscher}, J., {Huber}, D., {et~al.} 2010, \apjl, 713, L187

\bibitem[{{Hekker} {et~al.}(2011{\natexlab{b}}){Hekker}, {Gilliland},
  {Elsworth}, {Chaplin}, {De Ridder}, {Stello}, {Kallinger}, {Ibrahim},
  {Klaus}, \& {Li}}]{redgiants}
{Hekker}, S., {Gilliland}, R.~L., {Elsworth}, Y., {et~al.} 2011{\natexlab{b}},
  \mnras, 414, 2594

\bibitem[{{Hensberge} {et~al.}(2008){Hensberge}, {Iliji{\'c}}, \&
  {Torres}}]{2008A&A...482.1031H}
{Hensberge}, H., {Iliji{\'c}}, S., \& {Torres}, K.~B.~V. 2008, \aap, 482, 1031

\bibitem[{{Hilditch}(2001)}]{Hilditch01book}
{Hilditch}, R.~W. 2001, {An Introduction to Close Binary Stars} (Cambridge
  University Press, Cambridge, UK)

\bibitem[{{H{\o}g} {et~al.}(1997){H{\o}g}, {B{\"a}ssgen}, {Bastian}, {Egret},
  {Fabricius}, {Gro{\ss}mann}, {Halbwachs}, {Makarov}, {Perryman},
  {Schwekendiek}, {Wagner}, \& {Wicenec}}]{Hog+97aa}
{H{\o}g}, E., {B{\"a}ssgen}, G., {Bastian}, U., {et~al.} 1997, A\&A, 323, L57

\bibitem[{{Jester} {et~al.}(2005){Jester}, {Schneider}, {Richards}, {Green},
  {Schmidt}, {Hall}, {Strauss}, {Vanden Berk}, {Stoughton}, {Gunn}, \&
  {Others}}]{Jester+05aj}
{Jester}, S., {Schneider}, D.~P., {Richards}, G.~T., {et~al.} 2005, AJ, 130,
  873

\bibitem[{{Jiang} {et~al.}(2011){Jiang}, {Jiang}, {Christensen-Dalsgaard},
  {Bedding}, {Stello}, {Huber}, {Frandsen}, {Kjeldsen}, {Karoff}, {Mosser},
  {Demarque}, {Fanelli}, {Kinemuchi}, \& {Mullally}}]{chen}
{Jiang}, C., {Jiang}, B.~W., {Christensen-Dalsgaard}, J., {et~al.} 2011, \apj,
  742, 120

\bibitem[{{Kallinger} {et~al.}(2010){Kallinger}, {Weiss}, {Barban}, {Baudin},
  {Cameron}, {Carrier}, {De Ridder}, {Goupil}, {Gruberbauer}, {Hatzes},
  {Hekker}, {Samadi}, \& {Deleuil}}]{kallinger}
{Kallinger}, T., {Weiss}, W.~W., {Barban}, C., {et~al.} 2010, \aap, 509, A77+

\bibitem[{{Lehmann} {et~al.}(2011){Lehmann}, {Tkachenko}, {Semaan},
  {Guti{\'e}rrez-Soto}, {Smalley}, {Briquet}, {Shulyak}, {Tsymbal}, \& {De
  Cat}}]{lehmann}
{Lehmann}, H., {Tkachenko}, A., {Semaan}, T., {et~al.} 2011, \aap, 526, A124

\bibitem[{{Malkov} {et~al.}(2006){Malkov}, {Oblak}, {Snegireva}, \&
  {Torra}}]{ebcat}
{Malkov}, O.~Y., {Oblak}, E., {Snegireva}, E.~A., \& {Torra}, J. 2006, \aap,
  446, 785

\bibitem[{{Mazeh} \& {Zucker}(1992)}]{todcor1}
{Mazeh}, T. \& {Zucker}, S. 1992, in Astronomical Society of the Pacific
  Conference Series, Vol.~32, IAU Colloq. 135: Complementary Approaches to
  Double and Multiple Star Research, ed. H.~A. {McAlister} \& W.~I. {Hartkopf},
  164

\bibitem[{{Mazeh} \& {Zucker}(1994)}]{todcor3}
{Mazeh}, T. \& {Zucker}, S. 1994, \apss, 212, 349

\bibitem[{{Mosser} {et~al.}(2013){Mosser}, {Michel}, {Belkacem}, {Goupil},
  {Baglin}, {Barban}, {Provost}, {Samadi}, {Auvergne}, \& {Catala}}]{mosser13}
{Mosser}, B., {Michel}, E., {Belkacem}, K., {et~al.} 2013, \aap, 550, A126

\bibitem[{{Pasquini} {et~al.}(2011){Pasquini}, {Melo}, {Chavero}, {Dravins},
  {Ludwig}, {Bonifacio}, \& {de La Reza}}]{pasquini}
{Pasquini}, L., {Melo}, C., {Chavero}, C., {et~al.} 2011, \aap, 526, A127

\bibitem[{{Pr{\v s}a} {et~al.}(2011){Pr{\v s}a}, {Batalha}, {Slawson}, {Doyle},
  {Welsh}, {Orosz}, {Seager}, {Rucker}, {Mjaseth}, {Engle}, {Conroy},
  {Jenkins}, {Caldwell}, {Koch}, \& {Borucki}}]{ebcat1}
{Pr{\v s}a}, A., {Batalha}, N., {Slawson}, R.~W., {et~al.} 2011, \aj, 141, 83

\bibitem[{{Raskin} {et~al.}(2011){Raskin}, {van Winckel}, {Hensberge},
  {Jorissen}, {Lehmann}, {Waelkens}, {Avila}, {de Cuyper}, {Degroote},
  {Dubosson}, {Dumortier}, {Fr{\'e}mat}, {Laux}, {Michaud}, {Morren}, {Perez
  Padilla}, {Pessemier}, {Prins}, {Smolders}, {van Eck}, \& {Winkler}}]{hermes}
{Raskin}, G., {van Winckel}, H., {Hensberge}, H., {et~al.} 2011, \aap, 526, A69

\bibitem[{{Rucinski}(1999)}]{rucinski3}
{Rucinski}, S. 1999, in Astronomical Society of the Pacific Conference Series,
  Vol. 185, IAU Colloq. 170: Precise Stellar Radial Velocities, ed. J.~B.
  {Hearnshaw} \& C.~D. {Scarfe}, 82

\bibitem[{{Rucinski}(2002)}]{rucinski2}
{Rucinski}, S.~M. 2002, \aj, 124, 1746

\bibitem[{{Rucinski}(2004)}]{rucinski1}
{Rucinski}, S.~M. 2004, in IAU Symposium, Vol. 215, Stellar Rotation, ed.
  A.~{Maeder} \& P.~{Eenens}, 17

\bibitem[{{Salaris} \& {Girardi}(2002)}]{Salaris02}
{Salaris}, M. \& {Girardi}, L. 2002, \mnras, 337, 332

\bibitem[{{Sandquist} {et~al.}(2012){Sandquist}, {Mathieu}, {Brogaard},
  {Meibom}, {Geller}, {Orosz}, {Milliman}, {Jeffries}, {Brewer}, {Platais},
  {Grundahl}, {Bruntt}, {Frandsen}, \& {Stello}}]{sandquist}
{Sandquist}, E.~L., {Mathieu}, R.~D., {Brogaard}, K., {et~al.} 2012, ArXiv
  e-prints

\bibitem[{{Sing}(2010)}]{Sing10aa}
{Sing}, D.~K. 2010, A\&A, 510, A21

\bibitem[{{Slawson} {et~al.}(2011){Slawson}, {Pr{\v s}a}, {Welsh}, {Orosz},
  {Rucker}, {Batalha}, {Doyle}, {Engle}, {Conroy}, {Coughlin}, {Gregg},
  {Fetherolf}, {Short}, {Windmiller}, {Fabrycky}, {Howell}, {Jenkins}, {Uddin},
  {Mullally}, {Seader}, {Thompson}, {Sanderfer}, {Borucki}, \& {Koch}}]{ebcat2}
{Slawson}, R.~W., {Pr{\v s}a}, A., {Welsh}, W.~F., {et~al.} 2011, \aj, 142, 160

\bibitem[{{Southworth}(2008)}]{exolc}
{Southworth}, J. 2008, \mnras, 386, 1644

\bibitem[{{Southworth}(2011)}]{john}
{Southworth}, J. 2011, \mnras, 417, 2166

\bibitem[{{Southworth} {et~al.}(2007){Southworth}, {Bruntt}, \&
  {Buzasi}}]{Me++07aa}
{Southworth}, J., {Bruntt}, H., \& {Buzasi}, D.~L. 2007, A\&A, 467, 1215

\bibitem[{{Southworth} {et~al.}(2004){Southworth}, {Maxted}, \&
  {Smalley}}]{jktebop}
{Southworth}, J., {Maxted}, P.~F.~L., \& {Smalley}, B. 2004, \mnras, 351, 1277

\bibitem[{{Southworth} {et~al.}(2005{\natexlab{a}}){Southworth}, {Maxted}, \&
  {Smalley}}]{southworth}
{Southworth}, J., {Maxted}, P.~F.~L., \& {Smalley}, B. 2005{\natexlab{a}},
  \aap, 429, 645

\bibitem[{{Southworth} {et~al.}(2005{\natexlab{b}}){Southworth}, {Smalley},
  {Maxted}, {Claret}, \& {Etzel}}]{wwaur}
{Southworth}, J., {Smalley}, B., {Maxted}, P.~F.~L., {Claret}, A., \& {Etzel},
  P.~B. 2005{\natexlab{b}}, \mnras, 363, 529

\bibitem[{{Stello} {et~al.}(2013){Stello}, {Huber}, {Bedding}, {Benomar},
  {Bildsten}, {Elsworth}, {Gilliland}, {Mosser}, {Paxton}, \& {White}}]{stello}
{Stello}, D., {Huber}, D., {Bedding}, T.~R., {et~al.} 2013, \apjl, 765, L41

\bibitem[{{Stetson}(1987)}]{stet87}
{Stetson}, P.~B. 1987, \pasp, 99, 191

\bibitem[{{Thygesen} {et~al.}(2012{\natexlab{a}}){Thygesen}, {Frandsen},
  {Bruntt}, {Kallinger}, {Andersen}, {Elsworth}, {Hekker}, {Karoff}, {Stello},
  {Brogaard}, {Burke}, {Caldwell}, \& {Christiansen}}]{thygesen}
{Thygesen}, A.~O., {Frandsen}, S., {Bruntt}, H., {et~al.} 2012{\natexlab{a}},
  \aap, 543, A160

\bibitem[{{Thygesen} {et~al.}(2012{\natexlab{b}}){Thygesen}, {Frandsen},
  {Bruntt}, {Kallinger}, {Andersen}, {Elsworth}, {Hekker}, {Karoff}, {Stello},
  {Brogaard}, {Burke}, {Caldwell}, \& {Christiansen}}]{thygesen12}
{Thygesen}, A.~O., {Frandsen}, S., {Bruntt}, H., {et~al.} 2012{\natexlab{b}},
  \aap, 543, A160

\bibitem[{{Torres} {et~al.}(2010){Torres}, {Andersen}, \&
  {Gim{\'e}nez}}]{torres}
{Torres}, G., {Andersen}, J., \& {Gim{\'e}nez}, A. 2010, \aapr, 18, 67

\bibitem[{{Uytterhoeven} {et~al.}(2010){Uytterhoeven}, {Briquet}, {Bruntt}, {De
  Cat}, {Frandsen}, {Guti{\'e}rrez-Soto}, {Kiss}, {Kurtz}, {Marconi},
  {Molenda-{\.Z}akowicz}, {{\O}stensen}, {Randall}, {Southworth}, \&
  {Szab{\'o}}}]{uytterhoeven}
{Uytterhoeven}, K., {Briquet}, M., {Bruntt}, H., {et~al.} 2010, Astronomische
  Nachrichten, 331, 993

\bibitem[{{Zucker} \& {Mazeh}(1994)}]{todcor2}
{Zucker}, S. \& {Mazeh}, T. 1994, \apj, 420, 806

\end{thebibliography}




\end{document}